\documentclass[aps, prb, preprint]{revtex4-1}
\usepackage{bm, amsmath, amsfonts}
\usepackage{graphicx}
\usepackage{color}
\usepackage{braket}
\usepackage{hyperref}

\def\vector#1{\mbox{\boldmath$#1$}}
\def\bra#1{\langle\mbox{$#1$}\rvert}
\def\ket#1{\lvert\mbox{$#1$}\rangle}

\def\dx{\mathrm{d} x}
\def\dtau{\mathrm{d}\tau}

\def\zq{z^{(q)}}
\def\z2{z^{(2)}}
\def\J1J2{$J_1$-$J_2$}

\begin{document}


\title{Scaling of polarization amplitude in quantum many-body systems in one dimension}

\author{Ryohei Kobayashi}
\email{r.kobayashi@issp.u-tokyo.ac.jp}
\author{Yuya O. Nakagawa}
\author{Yoshiki Fukusumi}

\author{Masaki Oshikawa}
\affiliation{Institute for Solid State Physics, The University of Tokyo\\
Kashiwa, Chiba 277-8581, Japan.}

\date{\today}

\begin{abstract}
Resta proposed a definition of the electric polarization in
one-dimensional systems in terms of the ground-state expectation value of the large gauge transformation operator.
Vanishing of the expectation value in the thermodynamic limit implies that the system is a conductor.
We study Resta's polarization amplitude (expectation value) in the $S=1/2$ XXZ chain and its several generalizations, in the gapless conducting Tomonaga-Luttinger Liquid phase. We obtain an analytical expression in the lowest-order perturbation theory about the free fermion point (XY chain), and an exact result for the Haldane-Shastry model with long-range interactions. We also obtain numerical results, mostly using the exact diagonalization method.
We find that the amplitude exhibits a power-law scaling in the system size (chain length) and vanishes in the thermodynamic limit.
On the other hand, the exponent depends on the model even when the low-energy limit is described by the Tomonaga-Luttinger Liquid with the same Luttinger parameter. We find that a change in the exponent occurs when the Umklapp term(s) are eliminated, suggesting the importance of the Umklapp terms.
\end{abstract}


\pacs{*****}
\maketitle


\section{Introduction}
\label{sec:intro}

One of the most important problems in condensed matter physics is to identify electrical conduction properties of each material. As pointed out by Kohn~\cite{Kohn}, localization of electrons and the presence of a dielectric polarization density are two related essential features common to all insulating ground states of materials. As a consequence, the electric polarization could be utilized for the classification of conductor and insulator.
Based on several earlier studies~\cite{Zak,KSV1993,VKS1993,OrtizMartin},
Resta~\cite{Resta} proposed a compact definition of electric polarization, which can be naturally applied to interacting systems~\cite{RestaSorella} as well as to non-interacting electrons in one dimension.
In Resta's framework, the polarization for a ground state of a one-dimensional periodic lattice system with length $L$ is defined as $\mathrm{Im}{z}$, where
\begin{align}
z:=\bra{\psi_0}U\ket{\psi_0},
\label{Restadef}
\end{align}
which we call polarization amplitude.
Here $\ket{\psi_0}$ is a ground state, and
\begin{align}
U:=\exp{\left(\frac{2\pi i}{L}\sum_{j=1}^L j n_j\right)},
\label{eq.defU}
\end{align}
where $n_j$ is the particle number operator at site $j$.
The argument of the exponential in Eq.~\eqref{eq.defU}
is proportional to the center of mass of the particles, which is related to the polarization. The exponential form makes $U$ invariant under $j \to j+ L$ and naturally compatible with the periodic boundary condition. 
$U$ is nothing but the Lieb-Schultz-Mattis twist operator, or the large gauge transformation operator~\cite{LSM61,Oshikawa00,Nakamura02}.
Although it is interesting to consider extensions to higher dimensions, in this paper we focus on one-dimensional systems.

It was argued~\cite{Resta,RestaSorella} that the ampitude $z$ serves as a good indicator of electron localization in both non-interacting and interacting systems.
Intuitively, the polarization would be well-defined in an insulating phase when each electron is localized around nucleus, because one can define a local dipolar vector at each site, and many-body polarization is just defined by summing it over the whole system. On the other hand, in a conducting phase electrons are moving itinerantly and polarization would be ill-defined. Then it is natural to expect that the polarization amplitude $z$ can be an ``order parameter'' that distinguishes an insulating phase from conducting one. Resta conjectured that if the system is a conductor $z=0$ and an insulator $z\neq 0$ in the thermodynamic limit $L\rightarrow\infty$. 

It is easy to see this in free fermion systems. Since $U$ induces momentum shift by $2\pi/L$ for each particle, if one operates $U$ on a ground state of a gapless system, one particle is shifted from a Fermi point to another Fermi point, creating a particle-hole excitation.
This excited state is clearly orthogonal to the initial Fermi sea ground state, thus $z=0$.
On the other hand, if the system is a band insulator, $U \ket{\Psi_0}$ remains the ground state up to phase, and thus $|z| \to  1$ in the thermodynamic limit~\cite{Resta,WatanabeOshikawa}.



However, in the presence of a lattice translation symmetry $T$, one can immediately see that the simple criterion based on $z$ fails when the ground state is fractionally-filled. It holds that
\begin{align}
TUT^{-1}=e^{-2\pi i\nu}U,
\label{UT}
\end{align}
where $\nu$ is a filling factor, i.e., the number of electrons per a unit cell. 
It follows that
\begin{align}
\begin{split}
\bra{\psi_0}U\ket{\psi_0}&=\bra{\psi_0}T^{-1}TUT^{-1}T\ket{\psi_0}\\
&=e^{-2\pi i\nu}\bra{\psi_0}U\ket{\psi_0},
\end{split}
\end{align}
and thus $z=0$ when $\nu$ is not integer.
In fact, this observation is fundamental in the proof of the
celebrated Lieb-Schultz-Mattis (LSM) theorem~\cite{LSM61} and some of its generalizations~\cite{Oshikawa00}.

In a naive interpretation of $z$, $z=0$ would imply that the system is always conductor when it is fractionally-filled, but it is of course not true.
Indeed, the system can become a Mott insulator for any rational filling, if accompanied by a spontaneous breaking of the translation symmetry as required by the LSM theorem.
Based on this observation, Aligia and Ortiz~\cite{Aligia99} proposed using $U^q$ instead of $U$ when $\nu=p/q$ ($p$ and $q$ are coprime integers), i.e., they argued that the definition of polarization should be replaced by
\begin{align}
z^{(q)}:=\bra{\psi_0}U^q\ket{\psi_0},
\end{align}
so that the simple criterion $z^{(q)}\neq 0$ could be used to characterize insulators at any rational filling.
They have indeed confirmed that its consistency with Kohn's criterion for insulators based on the Drude weight~\cite{Kohn}.


The behavior of $z^{(q)}$ has been studied~\cite{Nakamura02, Todo02} in various insulating states, including the VBS state, the N\'{e}el ordered state, the gapped phase of bond-alternating Heisenberg chain, and the Mott insulating phase of the extended Hubbard model.
Analytical and numerical results confirmed that $z^{(q)} \neq 0$ in the thermodynamic limit.
However, comprehensive study of $z^{(q)}$ in gapless conducting phases of interacting particles has been lacking. The expected vanishing of $z^{(q)}$ in a conducting phase is already nontrivial for interacting systems. In a generic interacting system, $z^{(q)}$ does not vanish exactly in a finite-size system. Nevertheless, we expect that $z^{(q)}$ vanishes in the thermodynamic limit. If this is the case, we can ask how precisely $z^{(q)}$ vanishes as the system size increases, namely its scaling property.
We may hope that the scaling of $z^{(q)}$ characterizes various gapless conducting phases.

Toward this goal, in this paper, we study the polarization amplitude $z^{(q)}$ and its scaling in the $S=1/2$ XXZ chain
\begin{align}
H=J\sum_{j=1}^L \left({S}_j^x {S}_{j+1}^x + {S}_j^y{S}_{j+1}^y+\Delta {S}_j^z {S}_{j+1}^z \right),
\label{XXZham}
\end{align}
for $J>0$, with the periodic boundary condition
$S^{\alpha}_{L+1} \equiv S^{\alpha}_1$ ($\alpha=x,y,z$).
We will also study a few generalizations of the XXZ chain.

The $S=1/2$ XXZ chain~\eqref{XXZham} is one of the best studied models in quantum many-body problem. While it is often regarded as a model of one-dimensional quantum magnet, the same model can be understood as a model of hard-core bosons or fermions at half-filling ($\nu=1/2$) on a one-dimensional lattice, by identifying
\begin{equation}
 n_j = S^z_j + \frac{1}{2},
\end{equation}
as the particle number operator at site $j$.
In this way, we can naturally introduce the amplitude $z^{(q)}$ in the $S=1/2$ XXZ chain.

For $-1 < \Delta \leq 1$, the low-energy physics of
the XXZ chain is described as a Tomonaga-Luttinger liquid (TLL)\cite{Giamarchi}, which is a ubiquitous theory describing universal low-energy features of many one-dimensional quantum many-body systems. The TLL is nothing but the field theory in $1+1$ dimensions defined by the action
\begin{align} 
\label{free boson action} 
S_0[\phi]=\frac{1}{2\pi K}\int\dx\dtau\left[(\partial_\tau\phi)^2+(\partial_x\phi)^2\right],
\end{align}
where $\phi$ is a bosonic scalar field whose compactification radius is 1,
\begin{align}
\phi\sim\phi+2\pi,
\end{align}
and $K$ is a constant called the Luttinger parameter. The Luttinger
parameter $K$ determines various critical exponents. In fact, the TLL
represents a family of universality classes parametrized by $K$.
In the gapless critical regime $-1 < \Delta \leq 1$ of the XXZ chain for $J>0$,
the Luttinger parameter is exactly known\cite{Giamarchi} as
\begin{equation}
K = \frac{\pi}{2 \arccos{(-\Delta)}} .
\end{equation}
A few special values of $K$ are worth mentioning: $K=1$ describes free fermions, which corresponds to the XY chain ($\Delta=0$).
At $K=1/2$, the TLL acquires an enhanced SU(2) symmetry, which corresponds to the SU(2) symmetric antiferromagnetic Heisenberg chain ($\Delta=1$).

The $z$-component of spin operator, which corresponds to the particle number operator, is represented as 
\begin{equation}
S^{z}_{j}=\frac{1}{2\pi } \partial_{x} \phi+\left( -1\right)^{j }\cos \left( \phi\right),
\end{equation}
in the TLL theory.
This suggests that the polarization amplitude $\zq$ has the following field-theory representation:
\begin{align}
	\begin{split}
	\zq&=\left\langle \exp\left(\frac{2\pi qi}{L}\sum_{j=1}^L j\cdot S_j^z\right)\right\rangle \\
	&\overset{?}{=}\left\langle \exp\left(\frac{2\pi qi}{L}\sum_{j=1}^L j\cdot \left(\frac{1}{2\pi}\partial_x\phi(j)+(-1)^j\cos\phi(j)\right)\right)\right\rangle. \\
	\end{split}
\end{align}
This may be computed in a finite-size system by techniques in conformal field theory (CFT), as discussed in Appendix~\ref{app:TLL}. We find 
\begin{align}
\zq\overset{?}{\propto}\left(\frac{1}{L}\right)^{2q^2 K}.
\label{eq.zqCFT}
\end{align}
That is, the simple CFT calculation suggests that polarization amplitude $\zq$ decays as a power-law of the system size $L$, with the universal exponent solely determined by the Luttinger parameter $K$. This result is not only consistent with the expectation that $\zq$ vanishes in the thermodynamic limit for a gapless conducting phase, but also looks reasonable. However, as we will show later, it turns out that this naive field-theoretical result does not match the actual system size dependence observed analytically and numerically, even when the system is described by the TLL.

One of the possible sources of the discrepancy is that the free boson field theory~\eqref{free boson action} is only asymptotically exact in the low-energy limit. In general, the field-theory description of a given lattice model involves various irrelevant perturbations to the fixed point theory such as Eq.~\eqref{free boson action}. Even if these perturbations are ``irrelevant'' in the renormalization group sense, they can be essential in determining some physical quantities~\cite{FujimotoEggert}. In the case of the XXZ chain, so-called Umklapp terms exist as the irrelevant perturbations. With this in mind, we have studied several generalizations of the XXZ chain which correspond to suppression of the Umklapp term(s), analytically and numerically.

For all the models of gapless conducting phases we have studied, we find that the amplitude $\zq$ exhibits the power-law scaling
\begin{equation}
 \zq \propto \left( \frac{1}{L} \right)^{\beta(q)},
\label{eq.zqscaling}
\end{equation}
with the exponent $\beta(q) > 0$ depending on the model,
and vanishes in the thermodynamic limit.
This is in agreement with the original expectation. However, the value of the exponent $\beta(q)$ does not agree with the field-theory prediction~\eqref{eq.zqCFT}.
We also find that the power-law exponent $\beta$ changes substantially when the Umklapp term(s) is suppressed. This suggests the importance of the Umklapp term(s) on the amplitude $\zq$. However, we have not found a field-theory derivation of the observed ``non-universal'' results, even when the Umklapp terms are taken into account. At this moment, our findings present a challenging puzzle to the universal TLL description which is known to work well for virtually any other low-energy physical properties.


This paper is organized as follows. In Sec. \ref{sec:weak}, we first present analytical results of $\zq$ obtained by a perturbation theory for the XXZ chain near the free fermion (XY) point. We find a power-law scaling of $\zq$ in the system size $L$.
We also study the $J_1$-$J_2$ chain with the next-nearest-neighbor exchange interaction $J_2$ near the free fermion point in the perturbation theory. We again find the power-law scaling, but with a different exponent at a special value of $J_2/J_1$ (Gaussian point) where the leading Umklapp term vanishes.
Next, in Sec.~\ref{sec:HS}, we show the exact solution of $\zq$ for the ground state of Haldane-Shastry model, in which all the Umklapp terms are supposed to be absent. Contrary to our expectation, we still find a nontrivial power-law scaling which does not match the field-theory prediction. In Sec. \ref{sec:num}, we display numerical results on $z^{(q)}$ obtained by exact diagonalization for XXZ model and $J_1$-$J_2$ model at the Gaussian point. The observed power-law scalings generalize the results obtained by the perturbation theory to strongly interacting cases.
Furthermore, we also calculate $z^{(q)}$ numerically for the Gutzwiller-Jastrow wave function with the varying power, which generalizes the exact result on the Haldane-Shastry model.
Section~\ref{sec:discussion} is devoted to conclusion and discussion.



\section{Weak-coupling analysis}
\label{sec:weak}

As mentioned in the Introduction, 
the XXZ chain~\eqref{XXZham} can be regarded as a model
of (interacting) spinless fermion.
In particular, at $\Delta=0$, the model is often called as
the XY chain, which is exactly mapped to free fermions on
the one-dimensional lattice.

The $S^z S^z$ term with the coefficient $\Delta$
then represents the nearest-neighbor density-density interaction
of the fermions.
Even though the XXZ chain is exactly solvable for any value of $\Delta$
by the Bethe Ansatz~\cite{Baxter, Takahashi}, and its low-energy limit is known to be described
as the TLL, it is still useful to consider the system starting from
the free fermions and introduce $\Delta$ as a small perturbation.
This is particularly the case for our problem of the polarization amplitude $\zq$, which apparently defies a universal field-theory description.

\subsection{XY model}
\label{subsec:XY}

The $S=1/2$ XY chain, which corresponds to the special case of $\Delta=0$ of the XXZ chain~\eqref{XXZham}, can be mapped to the free fermion model
\begin{align}
H=-\frac{J}{2}\sum_{j=1}^L(c^\dagger_jc_{j+1}+\mathrm{h.c.}),
\label{freeham}
\end{align}
by the Jordan-Wigner transformation followed by a gauge transformation 
\begin{align}
\begin{split}
S_j^{+}&=(-1)^j\exp\left(i\pi\sum_{k=1}^{j-1}c^\dagger_kc_k\right)c_j^\dagger, \\
S_j^{-}&=c_j\exp\left(-i\pi\sum_{k=1}^{j-1}c^\dagger_kc_k\right)(-1)^j, \\
S_j^z&=c^\dagger_jc_j-\frac{1}{2}.
\label{JW}
\end{split}
\end{align}
For convenience in later discussions, we take $L=4N$ ($N$: integer). The ground state of the Hamiltonian (\ref{freeham}) is clearly the Fermi sea state
\begin{align}
\ket{\psi_0}=\prod_{-k_F<q<k_F}c^{\dagger}_q\ket{0},
\end{align}
where $k_F$ is a Fermi momentum: $k_F=\pi/2$, and the momentum $q$ takes values
\begin{align}
q=\frac{(2n+1)\pi}{L},
\end{align}
with $n=-N, -N+1,\dots, N-1$. To see that $\zq\equiv0$, we remark that $U$ induces the momentum shift of each fermion by $2\pi/L$:
\begin{align}
Uc^\dagger_qU^{-1}=c^\dagger_{q+2\pi/L},
\end{align}
where
\begin{align}
c^\dagger_q=\frac{1}{\sqrt{L}}\sum_{j=1}^L e^{iqj}c^\dagger_j.
\end{align}
Then, we can see that
\begin{align}
U^q\ket{\psi_0}\propto\prod_{n=1}^q c^{\dagger}_{k_F+\frac{(2n-1)\pi}{L}}c_{-k_F+\frac{(2n-1)\pi}{L}}\ket{\psi_0},
\end{align}
which is clearly orthogonal to the initial state $\ket{\psi_0}$. Therefore 
\begin{align}
\zq=\bra{\psi_0}U^q\ket{\psi_0}=0
\end{align}
for arbitrary $L$.


\subsection{XXZ model with a weak interaction}


The XXZ chain~\eqref{XXZham} is generally mapped to the model of interacting fermions
\begin{align}
H=-\frac{J}{2}\sum_{j=1}^L(c^\dagger_jc_{j+1}+\mathrm{h.c.})+J\Delta\sum_{j=1}^L\left(c^\dagger_jc_j-\frac{1}{2}\right)\left(c^\dagger_{j+1}c_{j+1}-\frac{1}{2}\right),
\end{align}
by the transformation (\ref{JW}).
When $\Delta$ is small, we can take the interaction as a perturbation, and the ground state $\ket{\psi}$ of (\ref{XXZham}) is expressed as 
\begin{align}
\ket{\psi}=\ket{\psi_0}+\sum_n\ket{\psi_n}\frac{1}{E_0-E_n}\bra{\psi_n}V\ket{\psi_0}
\end{align}
up to the 1st order perturbation, where $\ket{\psi_0}$ (resp. $\{\ket{\psi_n}\}$) is a ground state (resp. excited states) for $\Delta=0$, and $V$ is the interaction in $z$-direction:
\begin{align}
V=J\Delta \sum_{j=1}^L c^\dagger_jc_jc^\dagger_{j+1}c_{j+1}.
\end{align}
Then, the polarization becomes
\begin{align}
z^{(2)}=\sum_n\bra{\psi_0}V\ket{\psi_n}\frac{1}{E_0-E_n}\bra{\psi_n}U^2\ket{\psi_0}+\mathrm{c.c.}
\label{eq.z2.leading}
\end{align}
in the leading order of $\Delta$, where we used $\bra{\psi_0}U^2\ket{\psi_0}=0$.  $\bra{\psi_n}U^2\ket{\psi_0}$ takes nonzero value iff
\begin{align}
\ket{\psi_n}=U^2\ket{\psi_0}=c^\dagger_{k_F+\frac{\pi}{L}}c^\dagger_{k_F+\frac{3\pi}{L}}c_{-k_F+\frac{3\pi}{L}}c_{-k_F+\frac{\pi}{L}}\ket{\psi_0},
\end{align}
For this $\ket{\psi_n}$ the energy becomes $E_0-E_n=-2J(\sin\pi/L+\sin3\pi/L)$, and hence
\begin{align}
z^{(2)} & =
-\frac{1}{2J(\sin{\frac{\pi}{L}}+\sin{\frac{3\pi}{L}})}
\bra{\psi_0}Vc^\dagger_{k_F+\frac{\pi}{L}}c^\dagger_{k_F+\frac{3\pi}{L}}c_{-k_F+\frac{3\pi}{L}}c_{-k_F+\frac{\pi}{L}}\ket{\psi_0}+\mathrm{c.c.}\\
&=-\frac{\Delta}{\sin{\frac{\pi}{L}}+\sin{\frac{3\pi}{L}}}
\frac{1}{L}\left(-2+2\cos{\frac{2\pi}{L}}\right)
\approx\frac{\pi\Delta}{L^2},
\end{align}
thus we obtain the scaling law $z^{(2)}\propto1/L^2$ near $K=1$.
This indeed demonstrates that, $\zq$ can be non-vanishing in a finite-size system and shows a nontrivial power-law of the system size $L$, once the interaction among fermions is introduced. 

Similarly, we can obtain $z^{(2s)}$ in the leading,
$s$-th order of $\Delta$ as
\begin{align}
z^{(2s)} \sim \bra{\psi_0}(VR)^s U^{2s}\ket{\psi_0}+\mathrm{c.c.},
\label{pthorder}
\end{align}
where we introduced an operator $R$ as
\begin{align}
R=\sum_n\ket{\psi_n}\frac{1}{E_0-E_n}\bra{\psi_n}.
\end{align}
Evaluating Eq.~\eqref{pthorder} similarly to Eq.~\eqref{eq.z2.leading},
we find
\begin{align}
 z^{(2s)}\approx2\left(\frac{\Delta}{4}\right)^s\sum_{\sigma, \tau\in S_{2s}}\epsilon_\sigma\epsilon_\tau\prod_{j=1}^s \frac{(k_{\sigma(2j-1)}-k_{\sigma(2j)})(k_{\tau(2j-1)}-k_{\tau(2j)})}{L\left(\sum_{l=1}^{2j}(k_{\sigma(l)}+k_{\tau(l)})\right)},
\label{XXZqth}
\end{align}
where $k_j=(2j-1)\pi/L$, and $S_{2s}$ is the symmetric group of degree $2s$.
Each summand in (\ref{XXZqth}) is proportional to $1/L^{2s}$.
Hence we obtain, for $K \sim 1$,
\begin{equation}
 \beta(q) = q ,
\label{eq.beta.weak}
\end{equation}
for an even integer $q$.
However, it should be noted that, in the present analysis, we cannot rule out the possibility that the RHS of (\ref{XXZqth}) happens to vanish.
We will later confirm that the results of the perturbation theory obtained here are consistent with the numerical results on the XXZ chain.


\subsection{${J_1}$-${J_2}$ model}
\label{subsec:J1J2}

We have confirmed that, while $\zq$ exactly vanishes for the gapless free fermions, it is made finite (in a finite-size system) by the interaction.
In the field theory, the effects of the interaction may appear as (irrelevant) perturbation to the free boson field theory.
The XXZ chain has the U(1) symmetry generated by total magnetization $\sum_j S^z_j$. This symmetry, which we always keep in the present paper, forbids the perturbations of the form $\cos{(m\theta)}$ where $\theta$ is the dual field of $\phi$~\cite{Giamarchi}.
Moreover, the lattice translation symmetry is represented in TLL by $\phi \to \phi + \pi$, which forbids $\cos{\left((2n-1)\phi\right)}$.
Thus the effective action including the allowed perturbations reads
\begin{align}
 S[\phi]= S_0 + \sum_{n=1}^\infty g_{2n}
 \int dx d\tau \; \cos{(2n \phi)} + \ldots .
\end{align}
The vertex operators $\cos{(2n \phi)}$ represent the Umklapp processes of various orders, with the scaling dimensions $4 n^2 K$.
In the XXZ chain with $-1 < \Delta \leq 1$, $K \geq 1/2$ and thus the Umklapp operator is irrelevant.
As long as permitted by symmetries, we generically expect any perturbation to be non-vanishing: $g_{2n} \neq 0$ for any $n=1, 2,\dots$.
In order to see the importance of the Umklapp process, we can try to fine-tune the model to eliminate the leading Umklapp term $g_2$.


This can be indeed realized in the spin-1/2 ${J_1}$-${J_2}$ model
\begin{align}
H=J_1\sum_{j=1}^L \left({S}_j^x {S}_{j+1}^x + {S}_j^y{S}_{j+1}^y+\Delta {S}_j^z {S}_{j+1}^z \right)+J_2\sum_{j=1}^L \left({S}_j^x {S}_{j+2}^x + {S}_j^y{S}_{j+2}^y+\Delta {S}_j^z {S}_{j+2}^z \right)
\label{J1J2ham}
\end{align}
under the periodic boundary condition, which is transformed to a fermion system
\begin{align}
\begin{split}
H&=-\frac{J_1}{2}\sum_{j=1}^L(c^\dagger_jc_{j+1}+\mathrm{h.c.})+J_1\Delta\sum_{j=1}^L\left(c^\dagger_jc_j-\frac{1}{2}\right)\left(c^\dagger_{j+1}c_{j+1}-\frac{1}{2}\right)+\\
&\quad \  \frac{J_2}{2}\sum_{j=1}^L(c^\dagger_j(1-2c^\dagger_{j+1}c_{j+1})c_{j+2}+\mathrm{h.c.})+J_2\Delta\sum_{j=1}^L\left(c^\dagger_jc_j-\frac{1}{2}\right)\left(c^\dagger_{j+2}c_{j+2}-\frac{1}{2}\right),
\label{J1J2JW}
\end{split}
\end{align}
by the transformation (\ref{JW}).
This model has several phases~\cite{Nomura-Okamoto94, Hirata-Nomura00}
such as dimer phase or critical phase depending on $\Delta$ and $J_2$.


When we fix $\Delta$, the coefficient $g_2$ of the leading Umklapp term is non-zero for general values of $J_2$ in the critical phase. However, there is a special value $J_2$, which we denote $J_{2,G}(\Delta)$, where $g_2=0$ holds.
We call this point as the Gaussian point.
Vanishing of the leading Umklapp term $g_2=0$ in the field theory can be manifested, for example, 
in the absence of the logarithmic correction at the critical-dimer phase transition of the Heisenberg ($\Delta=1$) \J1J2 chain~\cite{Affleck89}.
As explained in Section \ref{subsec:XY}, $\zq \equiv 0$ for the free fermion system even when the system is finite ($L < \infty$).
From this viewpoint, the non-zero value of $\zq$ for finite system size $L$ may be attributed to the (irrelevant) Umklapp terms.
In order to investigate the effect of irrelevant terms, we also perform perturbation analysis for ${J_1}$-${J_2}$ model fine-tuned at the Gaussian point.

 In the case of $J_1\gg J_2\sim\Delta$, we can take the last three terms in (\ref{J1J2JW}) as perturbations, and in the lowest order of $\Delta$, $z^{(2)}$ becomes
\begin{align}
z^{(2)}=\sum_n\bra{\psi_0}V'\ket{\psi_n}\frac{1}{E_0-E_n}\bra{\psi_n}U^2\ket{\psi_0}+\mathrm{c.c.},
\end{align}
where
\begin{align}
V'=J_1\Delta\sum_{j=1}^Lc^\dagger_jc_jc^\dagger_{j+1}c_{j+1}-J_2\sum_{j=1}^L(c^\dagger_jc^\dagger_{j+1}c_{j+1}c_{j+2}+\mathrm{h.c.}),
\label{J1J2int}
\end{align}
which is 4-body interaction term of $O(\Delta)$ that appears in (\ref{J1J2JW}). The first term in (\ref{J1J2int}) corresponds to the $J_1$ $z$-direction interaction in the spin model, and we have already considered this type of contribution to $z^{(2)}$ in the previous subsection. The contribution of the second term in (\ref{J1J2int}) is
\begin{align}
&\frac{J_2}{2J_1(\sin{\frac{\pi}{L}}+\sin{\frac{3\pi}{L}})}
\bra{\psi_0}\left(\sum_{j=1}^L(c^\dagger_jc^\dagger_{j+1}c_{j+1}c_{j+2}+\mathrm{h.c.})\right)c^\dagger_{k_F+\frac{\pi}{L}}c^\dagger_{k_F+\frac{3\pi}{L}}c_{-k_F+\frac{3\pi}{L}}c_{-k_F+\frac{\pi}{L}}\ket{\psi_0}+\mathrm{c.c.}\\
=& \frac{2J_2}{J_1(\sin{\frac{\pi}{L}}+\sin{\frac{3\pi}{L}})}
\frac{1}{L}\left(\cos{\frac{6\pi}{L}}-2\cos{\frac{4\pi}{L}}+\cos{\frac{2\pi}{L}}\right)
 \approx-\frac{2\pi J_2}{J_1L^2}.
\end{align}
Therefore
\begin{align}
z^{(2)}=\frac{\pi}{L^2}\left(\Delta-\frac{2J_2}{J_1}\right)+O(1/L^4)
\end{align}
near $K=1$, and one can see that the scaling law becomes $z^{(2)}\propto1/L^4$ when $J_2/J_1=\Delta/2$, which corresponds to the Gaussian point of $J_1$-$J_2$ model in the limit of small $\lvert\Delta\rvert$. 

For $z^{(2s)}$, $s$-th order perturbation contributes to the leading term. Performing the similar calculation to the previous subsection, one can see that
\begin{align}
\begin{split}
z^{(2s)}&=\sum_{\sigma, \tau\in S_{2s}}\epsilon_\sigma\epsilon_\tau \prod_{j=1}^q \biggl( \frac{\Delta(e^{-ik_{\sigma(2j-1)}}-e^{-ik_{\sigma(2j)}})(e^{ik_{\tau(2j-1)}}-e^{ik_{\tau(2j)}})}{4L\left(\sum_{l=1}^{2j}(\sin k_{\sigma(l)}+\sin k_{\tau(l)})\right)} \\ 
& \qquad\qquad\qquad\qquad +\frac{\frac{J_2}{J_1}\left((e^{ik_{\sigma(2j-1)}}-e^{ik_{\sigma(2j)}})(e^{ik_{\tau(2j-1)}}-e^{ik_{\tau(2j)}})+\mathrm{c.c.}\right)}{4L\left(\sum_{l=1}^{2j}(\sin k_{\sigma(l)}+\sin k_{\tau(l)})\right)} \biggr)+\mathrm{c.c.},
\label{J1J2qth}
\end{split}
\end{align}
in the lowest order of $\Delta$. When $\Delta\neq2J_2/J_1$, the scaling is identical to the case of XXZ model: $z^{(2s)}\propto1/L^{2s}$ and
\begin{align}
z^{(2s)}=2\left(\frac{\Delta-\frac{2J_2}{J_1}}{4}\right)^s\sum_{\sigma, \tau\in S_{2s}}\epsilon_\sigma\epsilon_\tau\prod_{j=1}^s \frac{(k_{\sigma(2j-1)}-k_{\sigma(2j)})(k_{\tau(2j-1)}-k_{\tau(2j)})}{L\left(\sum_{l=1}^{2j}(k_{\sigma(l)}+k_{\tau(l)})\right)}
\label{J1J2L2q}
\end{align}
in the order of $1/L^{2s}$.  On the other hand, at the Gaussian point $\Delta=2J_2/J_1$, the scaling behavior drastically changes to $z^{(2s)}\propto1/L^{4s}$, and
\begin{align}
z^{(2s)}=2\left(\frac{\Delta}{8}\right)^s\sum_{\sigma, \tau\in S_{2s}}\epsilon_\sigma\epsilon_\tau\prod_{j=1}^s \frac{(k_{\sigma(2j-1)}^2-k_{\sigma(2j)}^2)(k_{\tau(2j-1)}^2-k_{\tau(2j)}^2)}{L\left(\sum_{l=1}^{2j}(k_{\sigma(l)}+k_{\tau(l)})\right)}
\label{J1J2L4q}
\end{align}
in the order of $1/L^{4s}$. 
Thus we find
\begin{equation}
 \beta(q) = 2q,
\label{eq.beta.J1J2.weak}
\end{equation}
for an even integer $q$, in the \J1J2 chain at the Gaussian point
near the XY limit ($K \sim 1$).
Thus we find that the exponent $\beta$ changes drastically
from Eq.~\eqref{eq.beta.weak} to Eq.~\eqref{eq.beta.J1J2.weak} 
by the fine-tuning of $J_2$ at the Gaussian point. 
This is consistent with our expectation that the Umklapp process has an important effect on the amplitude $\zq$. In fact, fine-tuning away the leading Umklapp term $g_2$ suppresses $\zq$ (by making the exponent $\beta$ larger),
as it is naturally expected.
However, we have not found a satisfactory field-theory derivation of the present observation. Moreover, we find a rather surprising result by eliminating the higher-order Umklapp terms, in the next section.


\section{Haldane-Shastry model (Gutzwiller-Jastrow wave function at $K=1/2$)}
\label{sec:HS}

We can eliminate the higher-order Umklapp terms $g_4, g_6, g_8, \ldots$ as well, by introducing and fine-tuning further neighbor couplings in the spin chain model. Although it is in principle possible to perform the fine-tuning successively, in practice it would be quite complicated.

Fortunately, however, the lattice realization of the ``fixed point'' theory~\eqref{free boson action} without the Umklapp terms is known as the Haldane-Shastry (HS) model with $1/r^2$-interaction~\cite{CalogeroSutherland,HaldaneShastry,Kato09}.
The Hamiltonian for a finite chain of length $L$ reads
\begin{align}
 H=\frac{J\pi^2}{L^2}\sum_{n<m}\frac{\vector{S}_{n}\cdot\vector{S}_{m}}{\sin^2(\pi(n-m)/L)} .
\label{HSham}
\end{align}
By identifying the down-spin state as an empty site (vacuum) and the up-spin state as a particle (magnon),
the ground state of this model is exactly given by the Gutzwiller-Jastrow wavefunction
as a function of the locations $x_i =1, 2,\ldots, L$ of the magnons
($i=1,2,\ldots, M$ is a label of magnons), as
\begin{align} \label{Jastrowstate}
 \tilde{\Psi}_G (x_1,\ldots,x_M) =
  \prod_i z_i^{-\frac{L(L-1)}{4K}}  \prod_{i<j} (z_i - z_j)^\frac{1}{K},
\end{align}
where
\begin{align}
 z_j = e^{2\pi i x_j /L} = e^{i \theta_j},
\end{align}
and
\begin{align}
 \theta_j = \frac{2 \pi x_j}{L} .
\end{align}
The Gutzwiller-Jastrow wavefunction~\eqref{Jastrowstate} for a general value of $K$ realizes the TLL with the Luttinger parameter $K$.
It has been found that the wavefunction~\eqref{Jastrowstate} have a large ($\sim 99.5\%$) overlap for $L=20$, with the ground state of XXZ chain~\cite{Cirac10} corresponds to the same Luttinger parameter $K$.
This type of wave function also appears in various important systems, such as the Laughlin state of the fractional quantum Hall effect (FQHE)~\cite{Laughlin83},
and the Calogero-Sutherland state of hard-core bosons~\cite{CalogeroSutherland}.
At the special value $K=1/2$, the TLL acquires the enhanced SU(2) symmetry, and Eq.~\eqref{Jastrowstate} is the exact ground state of the SU(2) symmetric Haldane-Shastry model~\eqref{HSham}.

As mentioned above, the Haldane-Shastry model and the associated Gutzwiller-Jastrow wavefunction ground state realizes the ``pure'' TLL in which all the Umklapp terms vanish.
If the non-vanishing amplitude $\zq \neq 0$ for finite system size in the XXZ chain were due to the Umklapp terms, we might expect that $\zq=0$ in the ground state of the Haldane-Shastry model. However, by an exact explicit calculation, we will show that $\zq \neq 0$ and that $\zq$ obeys a different scaling from that in the XXZ chain or at the Gaussian point of the $J_1$--$J_2$ model.


In order to find the normalized wavefunction 
\begin{equation}
 \Psi_G \propto \tilde{\Psi}_G,
\end{equation}
which satisfies
\begin{equation}
 \sum_{\{ x_j \}} | \Psi_G |^2 = 1,
\end{equation}
we need to obtain the norm
\begin{equation}
 \mathcal{N} = \sum_{\{ x_j \}} | \tilde{\Psi}_G |^2  .
\end{equation}
Although the evaluation of $\mathcal{N}$ is well known~\cite{Kato09},
for completeness we review the derivation in Appendix~\ref{app:norm}
since it also serves as a basis of the evaluation of the polarization.
The result reads
\begin{equation}
 \mathcal{N}= \left(\frac{L}{2}\right)^M\cdot M! (2M-1)!! = \left(\frac{L}{2}\right)^M\frac{L!}{2^M},
\label{eq.norm}
\end{equation}
where we used $L=2M$ in the ground state.


Now let us evaluate $z^{(2)}$ in the HS ground state.
\begin{equation}
z^{(2)} = 
\langle \Psi_G | U^2 | \Psi_G \rangle  =
\frac{1}{\mathcal{N}}
 \sum_{\{ x_j \}} \prod_j e^{2 i \theta_j} | \tilde{\Psi}_G (\{x_j\})|^2 .
\end{equation}
Following the logic in Appendix~\ref{app:norm},
\begin{align}
e^{2 i \theta_j} |\tilde{\Psi}_G|^2 = 
\left(\frac{1}{2} \right)^M
 \sum_P \epsilon_P &
 (p_{P2} - p_{P1}) e^{i(p_{P1} + p_{P2} + 2) \theta_1} 
\notag \\
& (p_{P4} - p_{P3}) e^{i(p_{P3} + p_{P4} + 2) \theta_2} 
\notag \\
& \ldots
 (p_{P(2M)}-p_{P(2M-1)}) e^{i(p_{P(2M-1)} + p_{P(2M)}+2) \theta_M} ,
\end{align}
where
\begin{equation}
 p_l = -M + \frac{1}{2}+ l = 
-M + \frac{1}{2}, -M + \frac{3}{2}, \ldots M-\frac{1}{2},
\end{equation}
and $P$ denotes the permutation. 

Now, each exponential is non-vanishing if and only if
$ p_{P(2j-1)}+ p_{P(2j)} = -2 $ or $L-2$. 
This condition is satisfied by $M$ distinct pairs
\begin{equation}
(p_1, p_{2M-2}), (p_2, p_{2M-3}), (p_3, p_{2M-2}), \ldots
(p_{M-1},p_M), (p_{2M-1},p_{2M}).
\end{equation}
Using the similar logic as in Appendix~\ref{app:norm}, the summation
over $\{ x_j\}$ gives
\begin{equation}
\left(\frac{L}{2}\right)^M\cdot M! (2M-3)!! 1 =\left(\frac{L}{2}\right)^M \frac{L \cdot (L-2)!}{2^M}
\end{equation}
Dividing by the norm~\eqref{eq.norm}, we find
\begin{equation}
 \frac{1}{L-1}.
\end{equation}
For a large $L$, this reduces to the simple power law $L^{-1}$.

It is straightforward to extend this result to the expectation value of
$U^q$ for general $q$.
We find, for an even integer $q$,
\begin{align}
 \zq_{K=1/2, \mathrm{Jastrow}}=\prod_{j=1}^{q/2}\frac{2j-1}{L-2j+1} \sim
 \frac{1}{L^{q/2}},
\end{align}
whereas $\zq=0$ for an odd integer $q$ as required.
Thus we find
\begin{equation}
\beta(q) = \frac{q}{2}
\label{HSresults} 
\end{equation}
for even integer $q$ in the Haldane-Shastry model.

This result is remarkable in several respects.
First, a compact analytical expression which is exact even for a finite size $L$ is obtained for the nontrivial polarization amplitude $z^{(q)}$ in the strongly interacting many-body system. The result shows a simple power-law scaling, which is consistent with a general expectation for gapless conductors. However, the non-vanishing (for a finite size) $z^{(q)}$ in the complete absence of the Umklapp terms is against the simple picture that a non-vanishing amplitude $z^{(q)}$ is induced by the Umklapp terms.
On the other hand, since the Haldane-Shastry model is considered to be an ideal realization of the TLL without the Umklapp terms, we might expect that the prediction~\eqref{eq.zqCFT} based on the free boson field theory would apply. However, the actual exact results~\eqref{HSresults} on the Haldane-Shastry model does not agree with Eq.~\eqref{eq.zqCFT}.
In fact, at this point we do not have a field-theory understanding of the exact results~\eqref{HSresults}.


\section{Numerical Approach}
\label{sec:num}

In the previous sections, we have studied the amplitude $z^{(q)}$ analytically. However as in the case of most physical quantities, analytical results are available only for limited cases.
In order to study the amplitude $z^{(q)}$ and its scaling in a wider class of models, in this section, we employ numerical methods.
We obtain the amplitude $z^{(q)}$ in the ground states of the standard XXZ chain~\eqref{XXZham} and of the \J1J2 XXZ model at the Gaussian point where the leading Umklapp term is eliminated, by numerical exact diagonalization.
The most severe drawback of the numerical exact diagonalization is the limitation to small system sizes. However, in most of the cases we studied, the numerical exact diagonalization of finite chains up to $L=26$ sites was enough to find a power-law scaling of $z^{(q)}$ in $L$. Furthermore, we study $z^{(q)}$ numerically for the Gutzwiller-Jastrow
wave function~\eqref{Jastrowstate} at generic values of $K$ for which we have not found an exact result by the combinatorial method as in Sec.~\ref{sec:HS}.



\subsection{XXZ chain}

\begin{figure}
 \includegraphics[width = 8cm]{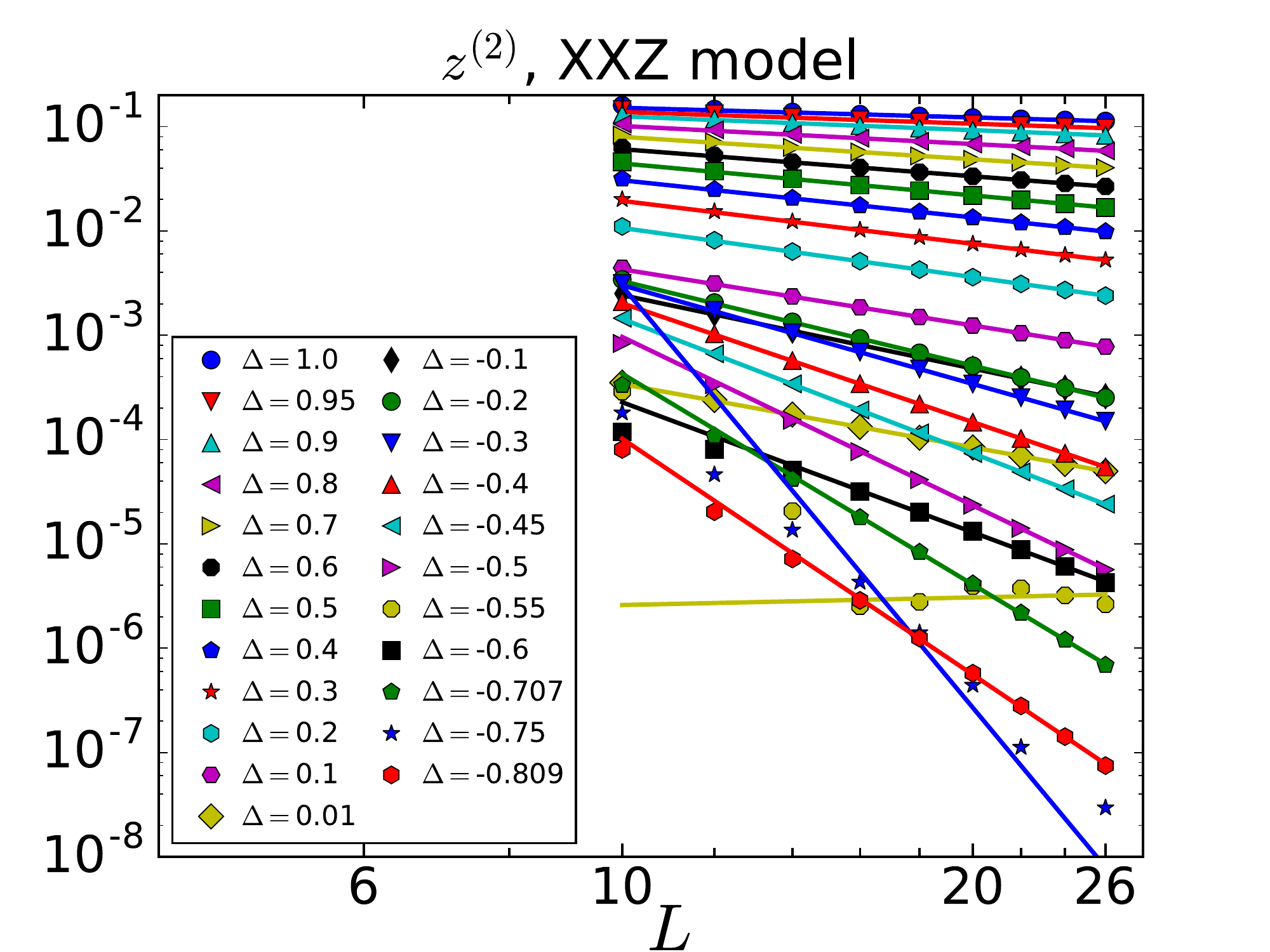}
 \includegraphics[width = 8cm]{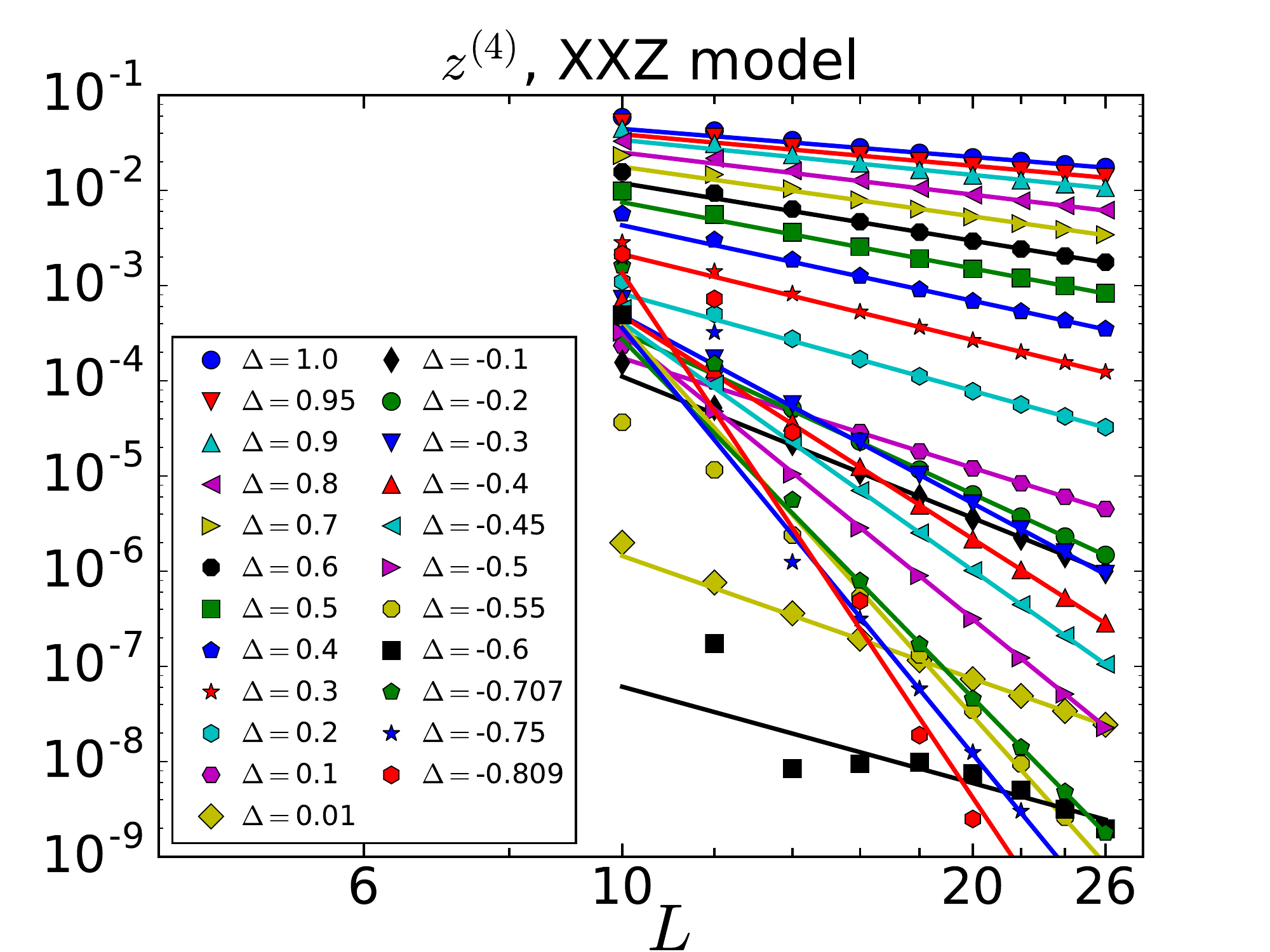}
 \includegraphics[width = 8cm]{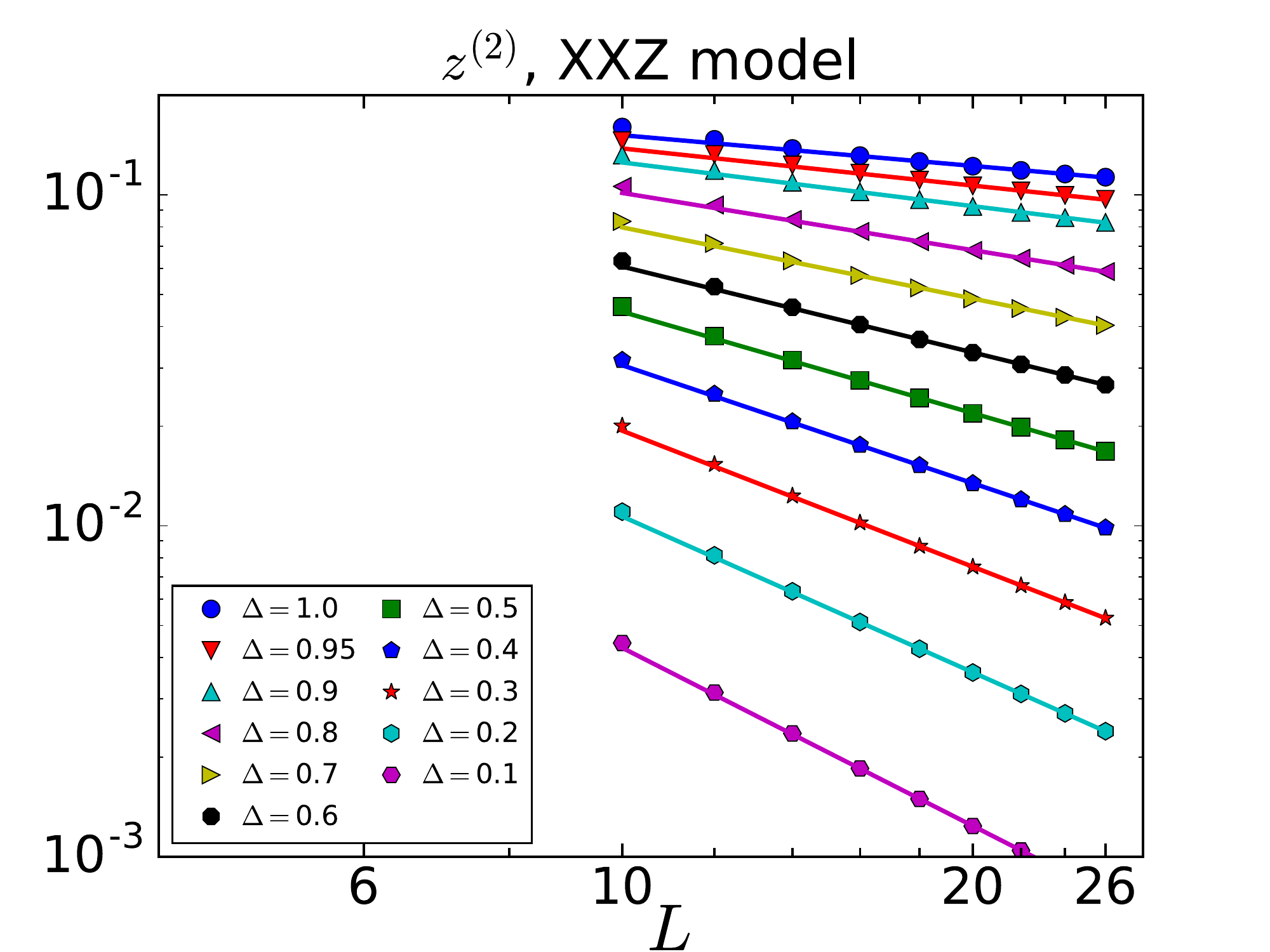}
 \includegraphics[width = 8cm]{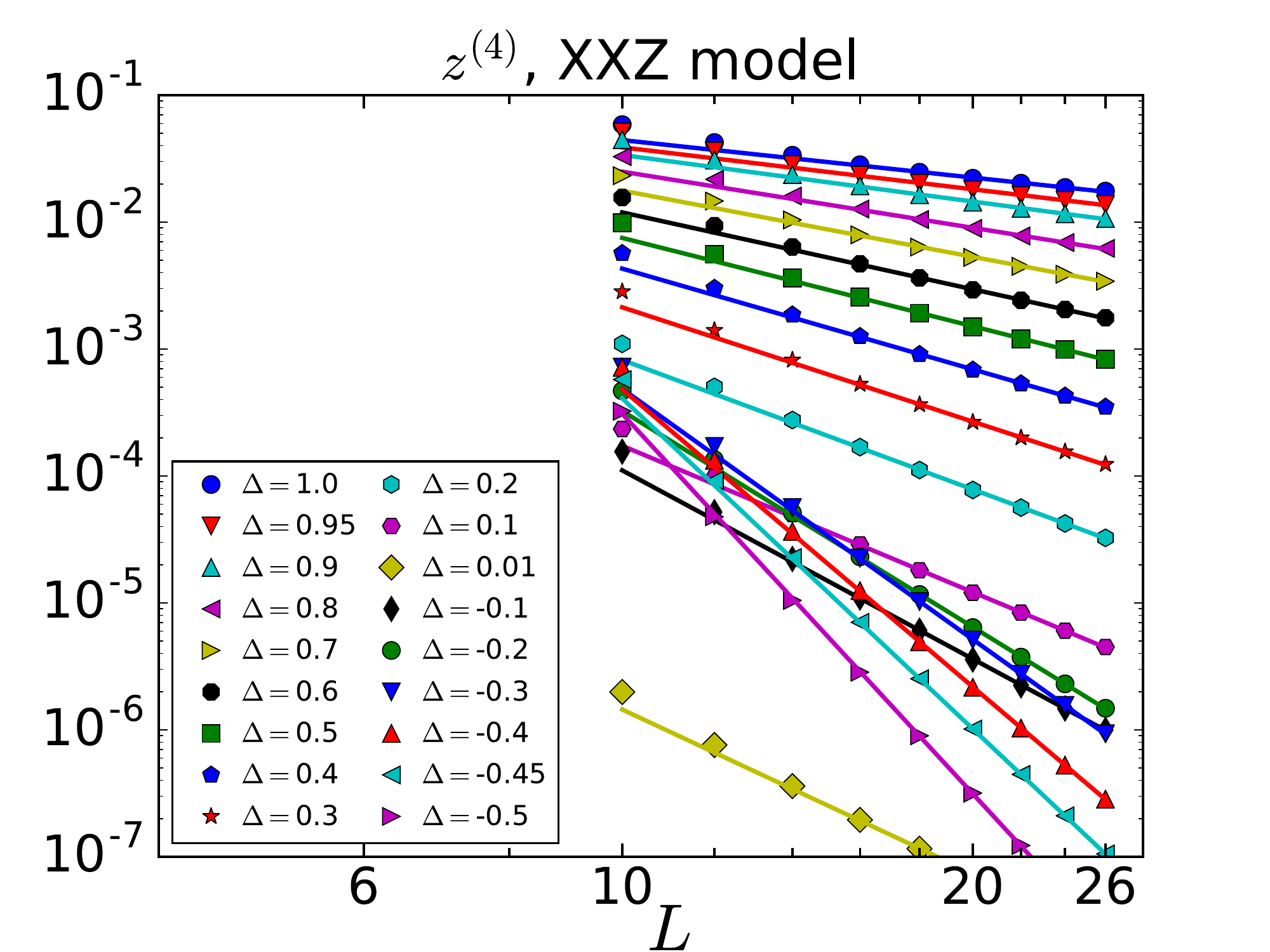}
 \includegraphics[width = 8cm]{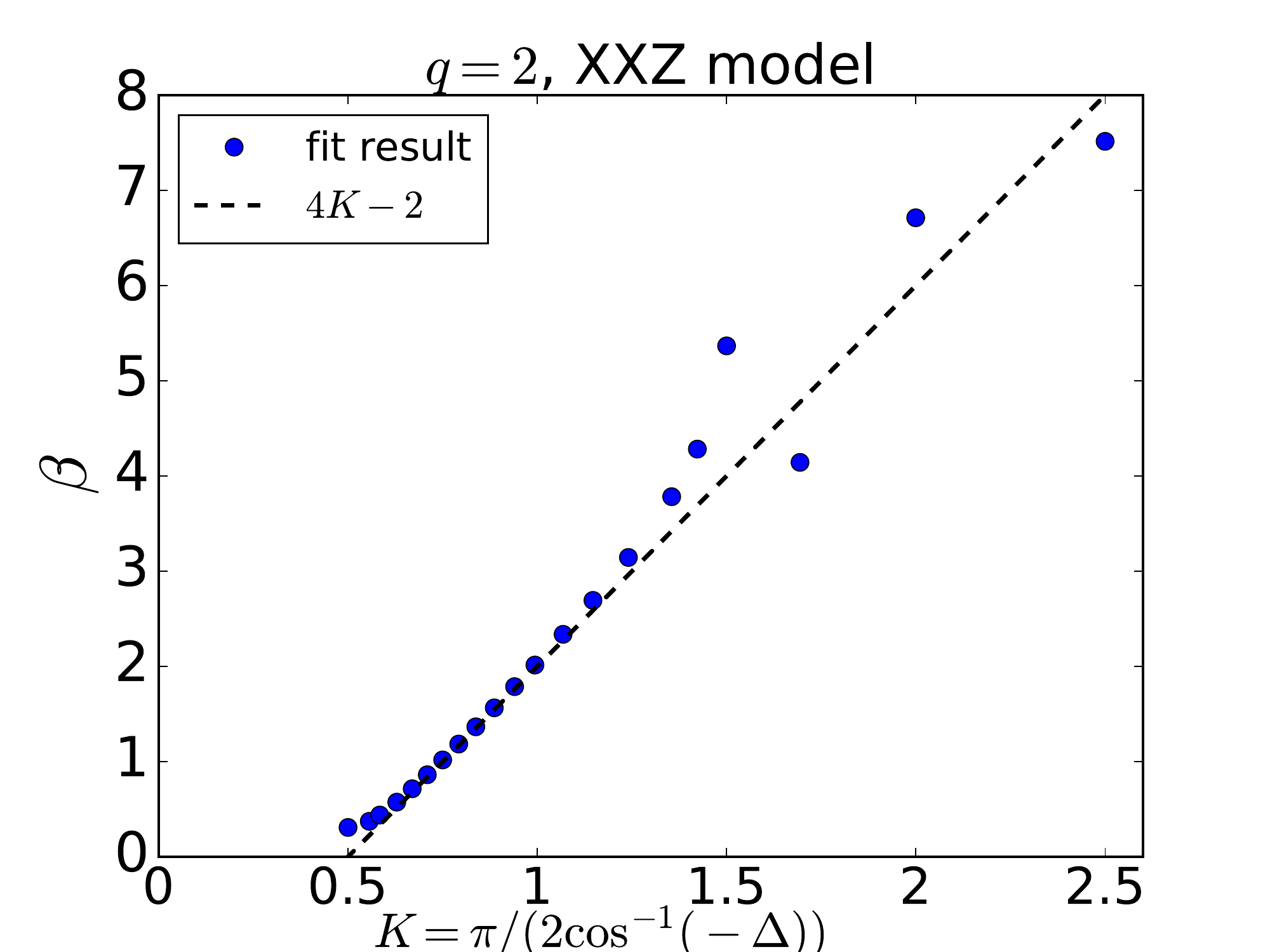}
 \includegraphics[width = 8cm]{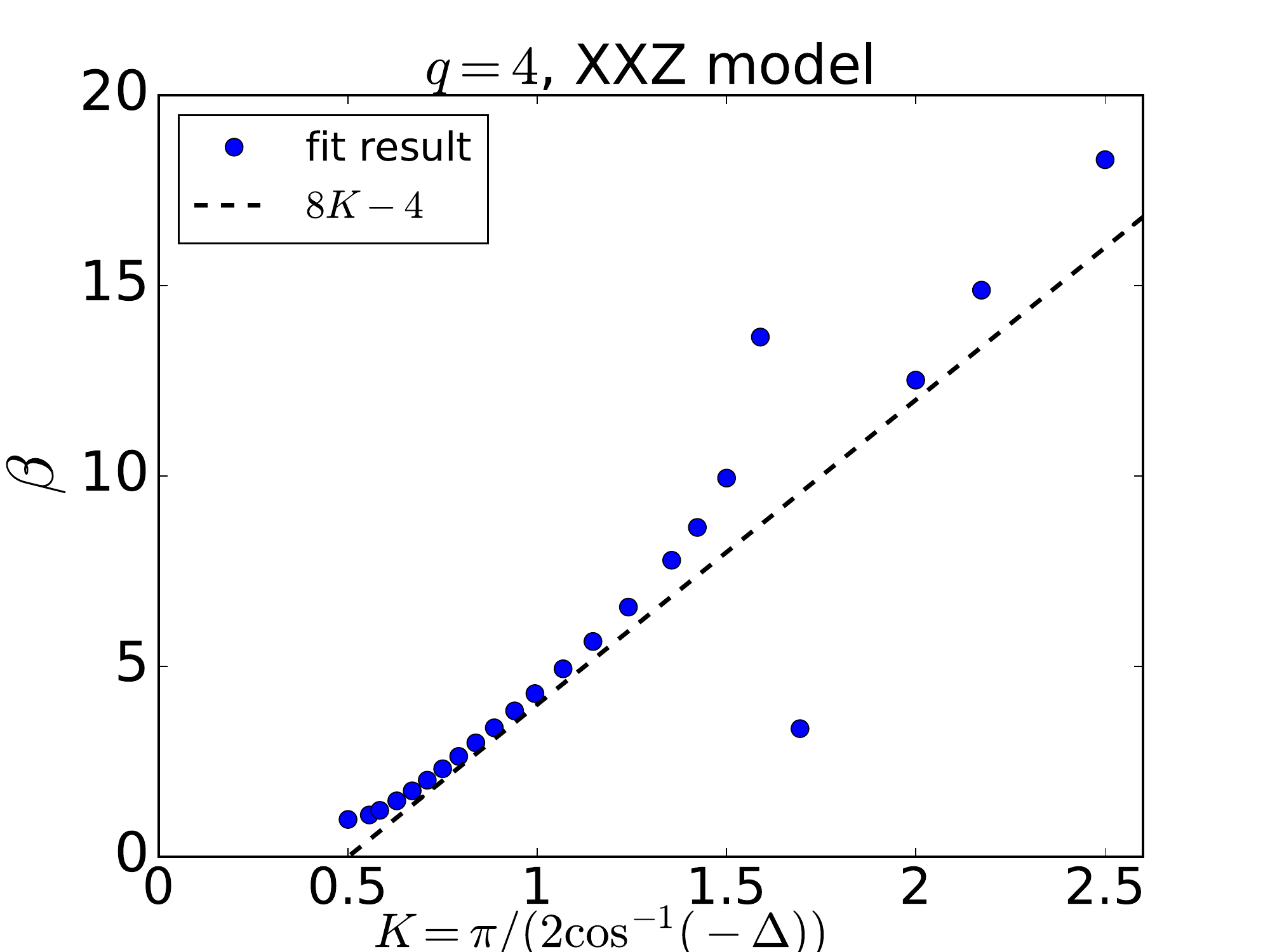}
 \caption{
 (top)
 Numerical results of $\z2$ (left) and $z^{(4)}$ (right) for the ground state of the XXZ chain~\eqref{XXZham} with the system size $L$ up to $L=26$.
 The dots are the numerical data and the lines are numerical fits by a simple power-law $f(L) = a / L^\beta$ where $a$ and $\beta$ are fitting parameters.
 (middle) The closeups of the top panels.  
 (bottom) The power-law exponent $\beta$ obtained from the fitting.
  }
 \label{XXZ_numerics}
\end{figure}

First let us present the results of numerical exact diagonalization of the standard XXZ chain~\eqref{XXZham} up to the system size $L=26$.
In the top left and middle left panels of Fig.~\ref{XXZ_numerics}, we present $\z2$ in the ground state of \eqref{XXZham}.
The power-law decay of $\z2$ with $L$ is clearly visible for $ -0.5 \le \Delta \le 1$.
However, for $ \Delta <  -0.5 $, the power-law scaling is less clear.
Especially, the data of $\Delta = -0.55$ do not show any power-law decay within the system size we can reach ($L=26$).
This seemingly strange change of the behavior across $\Delta = -0.5$ can also be seen in the left bottom panel of Fig.~\ref{XXZ_numerics}, where the power-law exponents $\beta$ estimated from the fitting of the data of $\z2$ are plotted.
Around $K=1.5$, or $\Delta=-0.5$, the exponent $\beta$ exhibits non-systematic behavior
(we note that the data corresponding to $\Delta=-0.55$ is not plotted in the figure.).
We see that the overall behavior of $\beta$ is explained by $\beta = 4K-2$, especially for $K<1$.
In the panels of the right column of Fig.~\ref{XXZ_numerics}, we present the data of $z^{(4)}$.
The behaviors are qualitatively the same as those of $\z2$ and the exponent of the power-law $\beta$
might be described by $\beta = 8K-4$, which is twice of the value of the $q=2$.
Thus we can conjecture
\begin{equation}
 \beta(q) = q (2K-1),
\label{eq.beta.XXZ}
\end{equation}
for an even integer $q$, in the XXZ chain with $K \lesssim 1.5$.

We have several comments in order.
First, the present result is consistent with Eq.~\eqref{eq.beta.weak} obtained by the weak-coupling perturbation theory for $K \sim 1$.
Second, although the relation $\beta = q(2K-1)$ seems to hold well both for $q=2$ and $q=4$ for $K \lesssim 1.5$,
the exponent $\beta$ deviate from this relation around the Heisenberg point $\Delta=1$ or $K=1/2$
(numerical fit suggests that the exponent $\beta$ is 0.25 for $q=2$ and 0.5 for $q=4$). This deviation may be attributed to the logarithmic correction caused by the marginally irrelevant interaction $g_2$ for $K=1/2$.
As we will see in the next subsection, we do not find such a deviation in the case of \J1J2 chain at the Gaussian point where the leading Umklapp term $g_2$ is absent. This is consistent with the above reasoning.
Finally, Eq.~\eqref{eq.beta.XXZ} seems to break down for $K \gtrsim 1.5$.
We must be cautious in drawing a conclusion since
the exponent $\beta$ for $K \gtrsim 1.5$ obtained in the numerical calculations might not be reliable because we do not reach large enough system size $L$ to see clear power-law behaviors.
Nevertheless, since we observe similar departure of $\beta(q)$ from a simple linear function of $K$ at $K \sim 1.5$ also in other models (see the following subsections), it is tempting to identify some kind of transition or crossover at $K \sim 1.5$.
However, at present we do not have any theoretical understanding for this.

\subsection{ \J1J2 XXZ chain tuned at the Gaussian point}

In order to study the effect of the leading Umklapp term, next we study the $J_1$-$J_2$ model as introduced in Section~\ref{subsec:J1J2}.



\begin{figure}
 \includegraphics[width = 8cm]{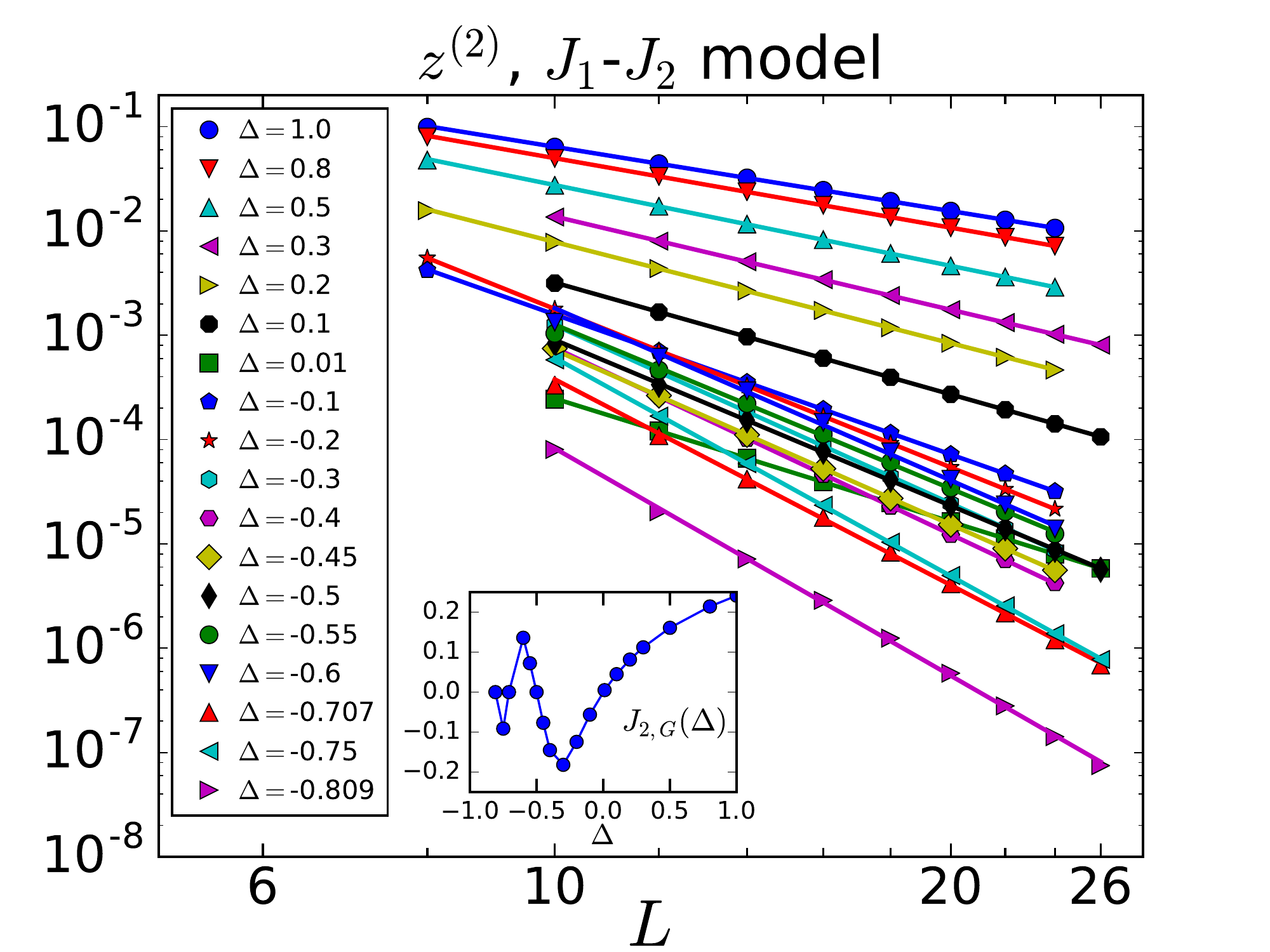}
 \includegraphics[width = 8cm]{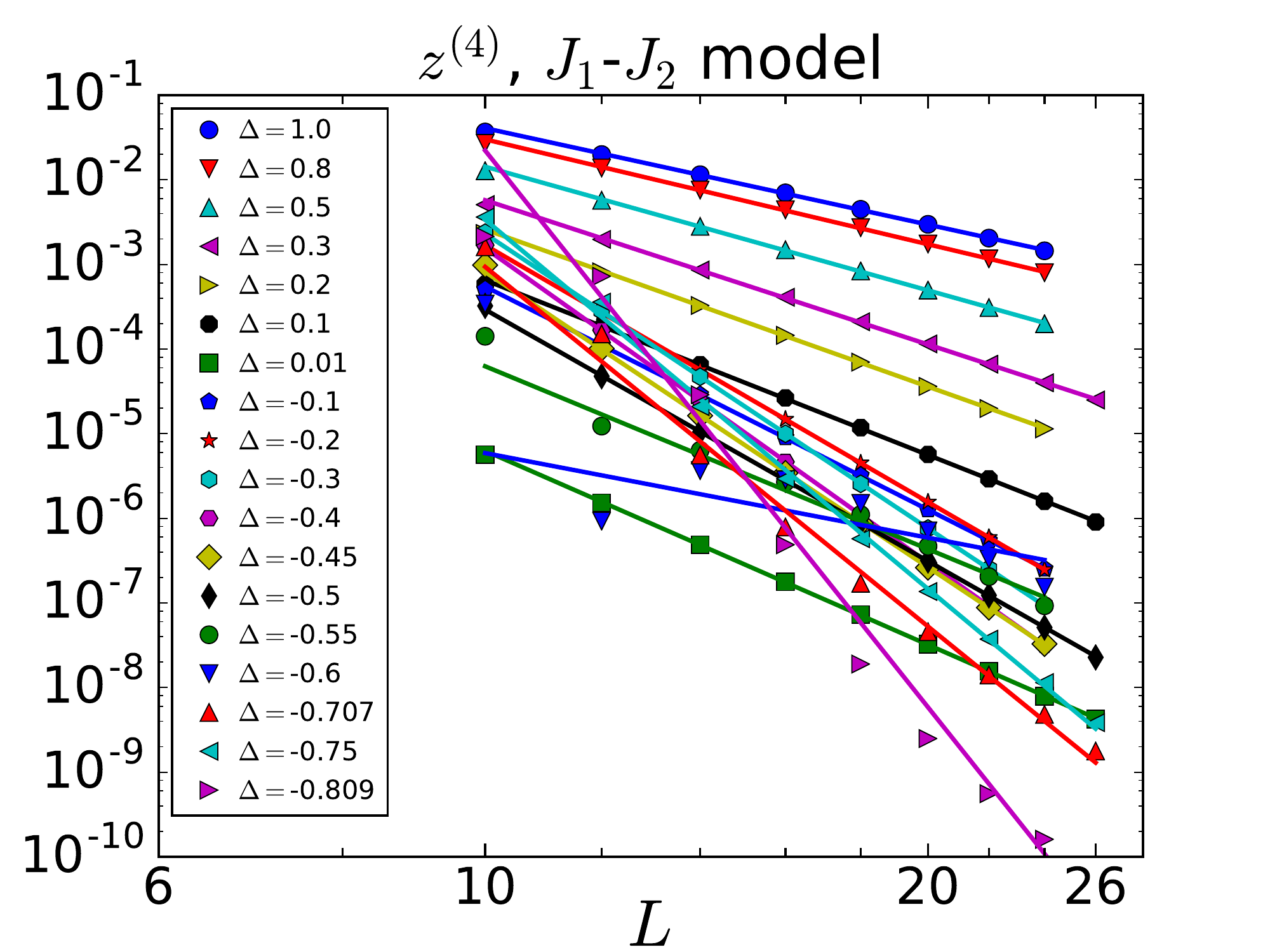}
 \includegraphics[width = 8cm]{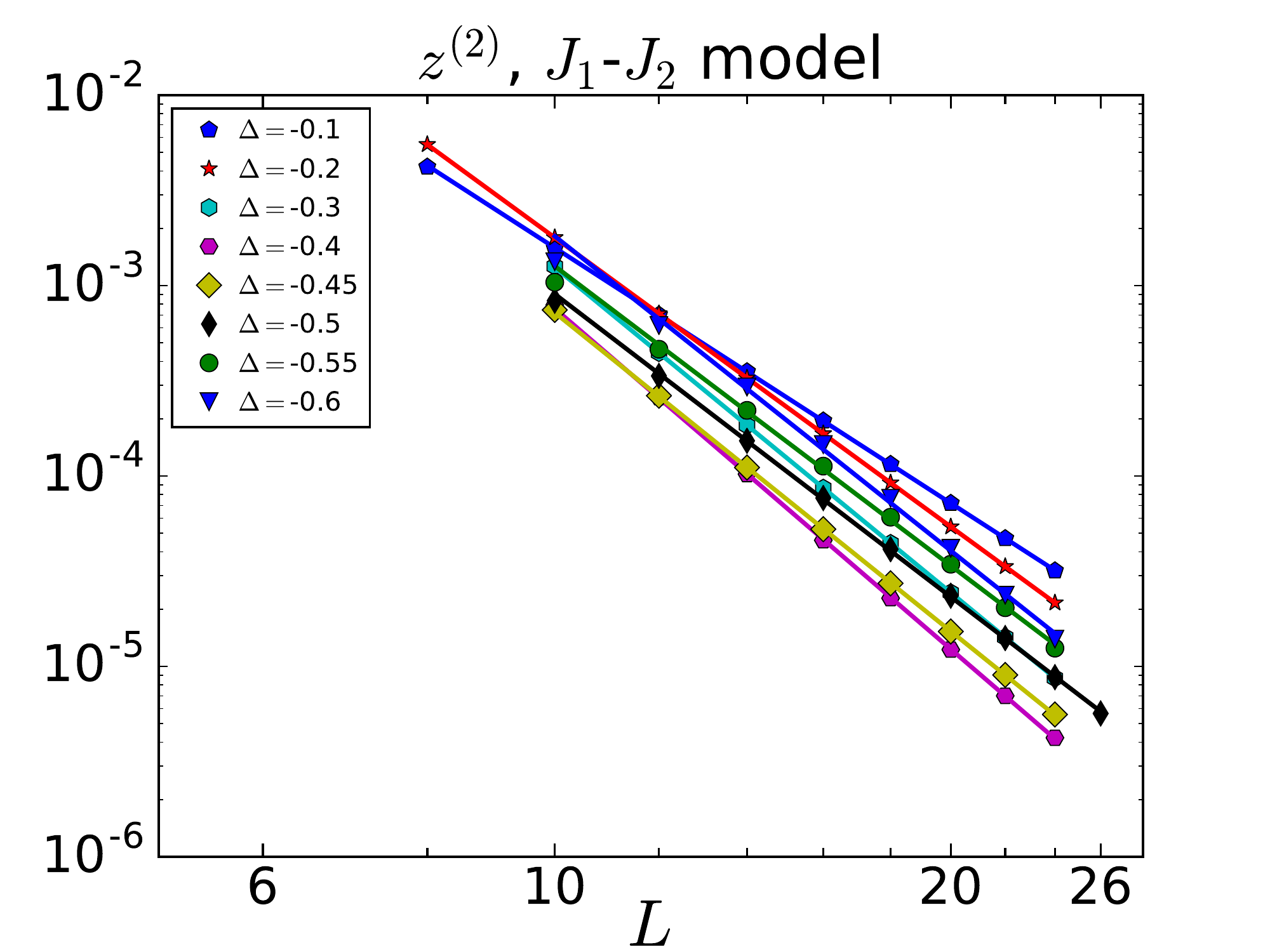}
 \includegraphics[width = 8cm]{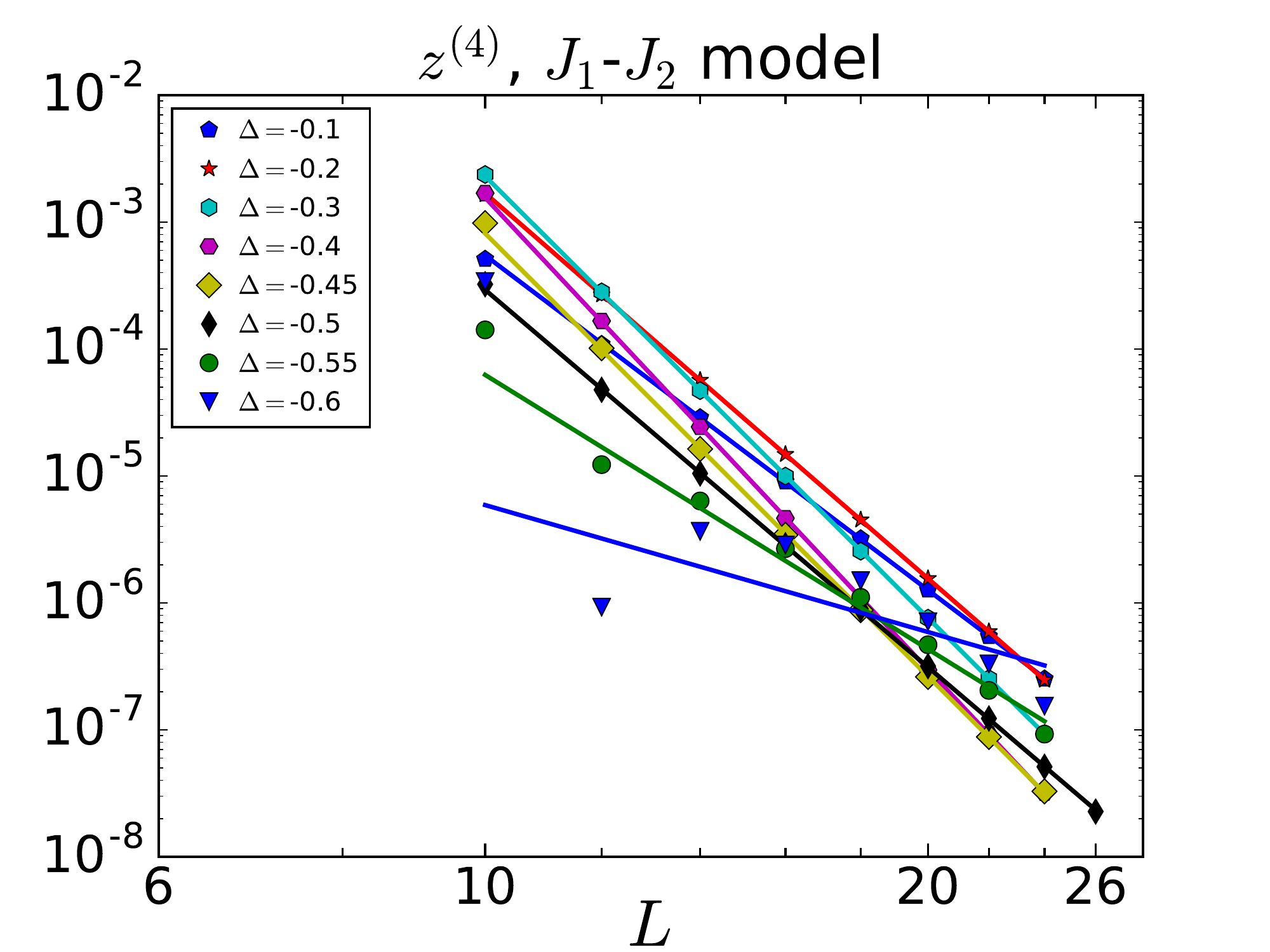}
 \includegraphics[width = 8cm]{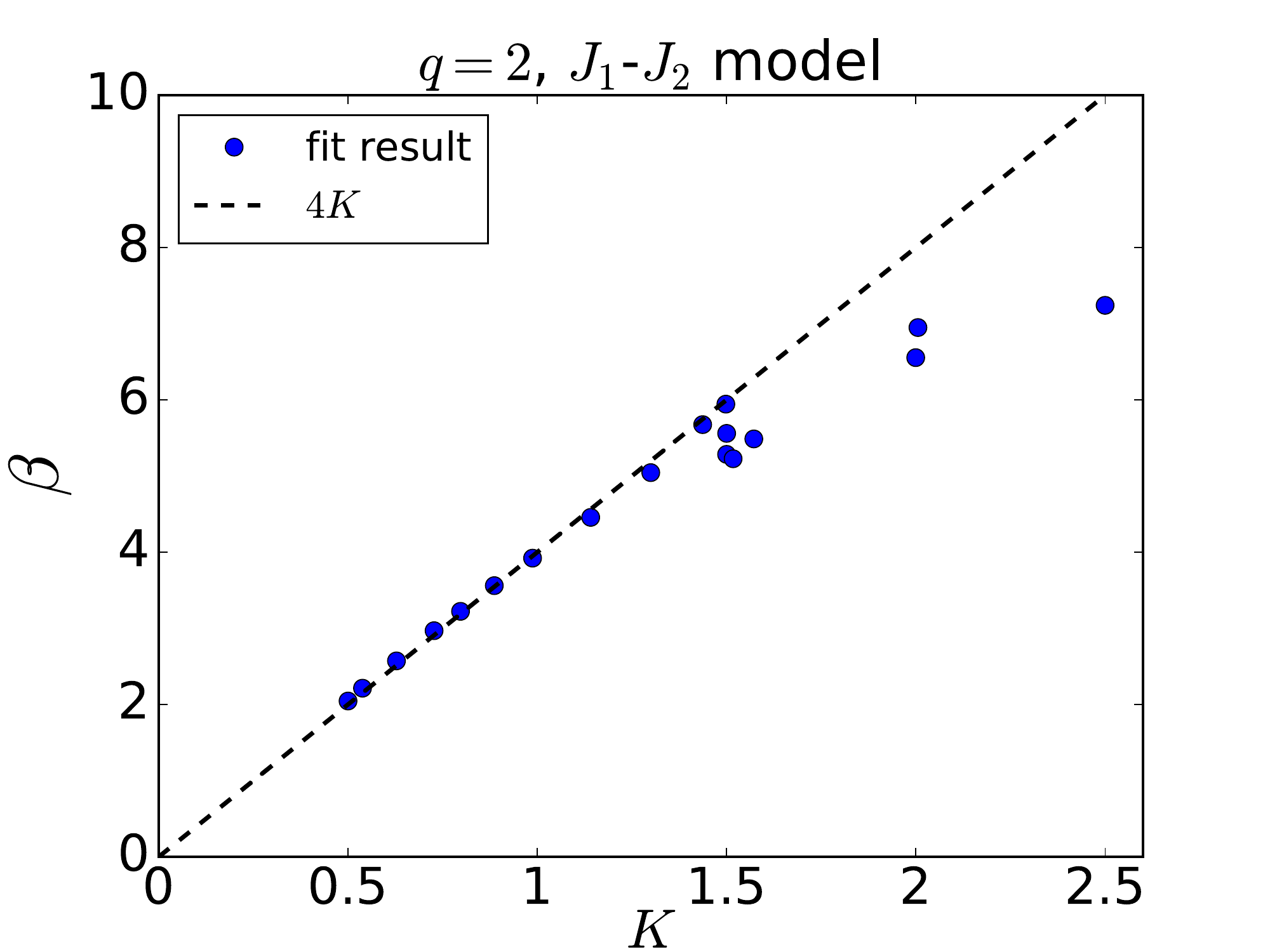}
 \includegraphics[width = 8cm]{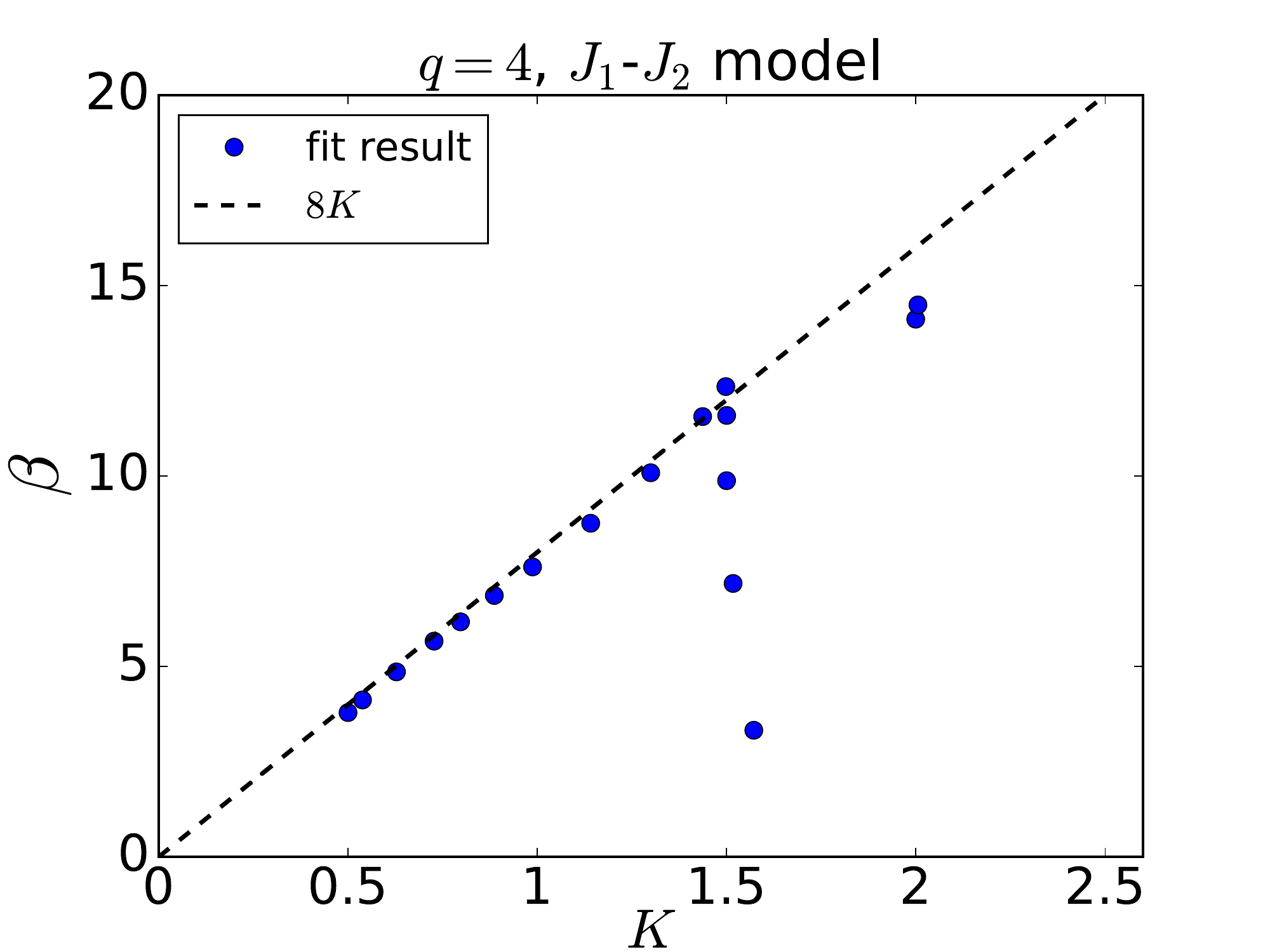}
 \caption{
 (top)
 Numerical results of $\z2$ (left) and $z^{(4)}$ (right) for the ground state of the \J1J2 XXZ model~\eqref{J1J2ham} at the Gaussian fixed point with the system size $L$.
 The dots are the numerical data and the lines are numerical fits by a simple power-law $f(L) = a / L^\beta$ where $a$ and $\beta$ are fitting parameters.
 (middle) The closeups of the top panels.  
 (bottom) The power-law exponent $\beta$ obtained from the fitting.
  }
 \label{J1J2_numerics}
\end{figure}

First we need to identify the Gaussian point $J_{2,G}(\Delta)$ where the leading Umklapp term vanishes ($g_2=0$).
In Section~\ref{subsec:J1J2} it was done analytically in the lowest order of the perturbation theory in the interaction $\Delta$. For generic values of $\Delta$, no explicit formula for $J_{2,G}(\Delta)$ is available. Therefore, we have to determine $J_{2,G}(\Delta)$ numerically.
This was done with the level spectroscopy method~\cite{Nomura-Okamoto94, Nomura95}.
Setting $J_2$ to $J_{2,G}(\Delta)$ thus obtained, we numerically obtain the amplitude $\zq$, as we did for the standard XXZ chain. Furthermore, we also determine the Luttinger parameter $K$ by evaluating the energy-level spacing of the system. More technical details on the numerical calculations are presented in Appendix~\ref{app: J1J2 Gaussian}.
The top left and middle left panels of Fig.~\ref{J1J2_numerics} shows the results of $\z2$ obtained by exact diagonalization.
$\z2$ exhibits a clear power-law decay for all values $\Delta$ even for $\Delta < -0.5$ in contrast to the XXZ chain~\eqref{XXZham} in the previous subsection.
In the inset of the top left panel, the value of $J_{2,G}(\Delta)$ is also shown.
As for the power-law exponent $\beta$, we numerically find that $\beta = 4K$ explains the data well for $K \lesssim1.5$ (the bottom left panel of Fig.~\ref{J1J2_numerics}).
We also show the numerical results for $z^{(4)}$ in  the panels in the right column of Fig.~\ref{J1J2_numerics}, which imply $\beta=8K$ for $K \lesssim 1.5$.
Thus we conjecture
\begin{equation}
 \beta(q) = 2qK,
\end{equation}
for an even integer $q$, in the \J1J2 chain at the Gaussian point with
$K \lesssim 1.5.$
Again this is consistent with the weak-coupling result~\eqref{eq.beta.J1J2.weak}
near the XY point, for $K \sim 1$.
Remarkably, a steep change or possible discontinuity of $\beta$ is observed
at $K\sim 1.5$, as in the case of the XXZ chain.
Again we do not have a theoretical understanding for this phenomenon at $K \sim 1.5$.

\subsection{Gutzwiller-Jasrtow wave function}

Finally, we study the polarization amplitude $\zq$ in the Gutzwiller-Jastrow wave function~\eqref{Jastrowstate}.
Exact results were presented for $K=1/2$, which corresponds to the SU(2) symmetric Haldane-Shastry model, in Sec.~\ref{sec:HS}.
Here we study the same wave function but at different values of $K$.
For generic values of $K$, we have not found exact results on $\zq$ and thus we need to evaluate $\zq$ numerically.

\begin{figure*}
 \includegraphics[width = 8cm]{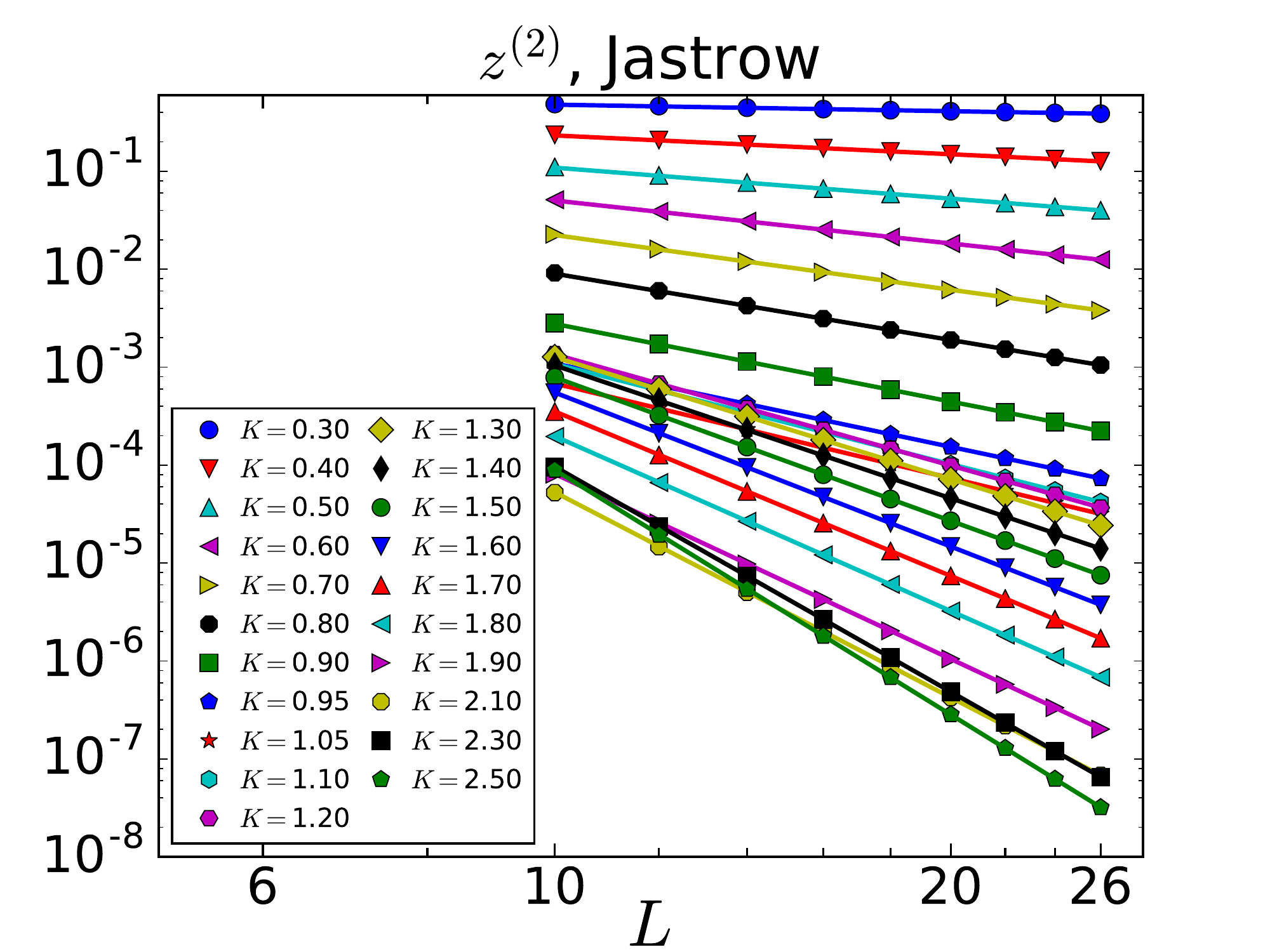}
 \includegraphics[width = 8cm]{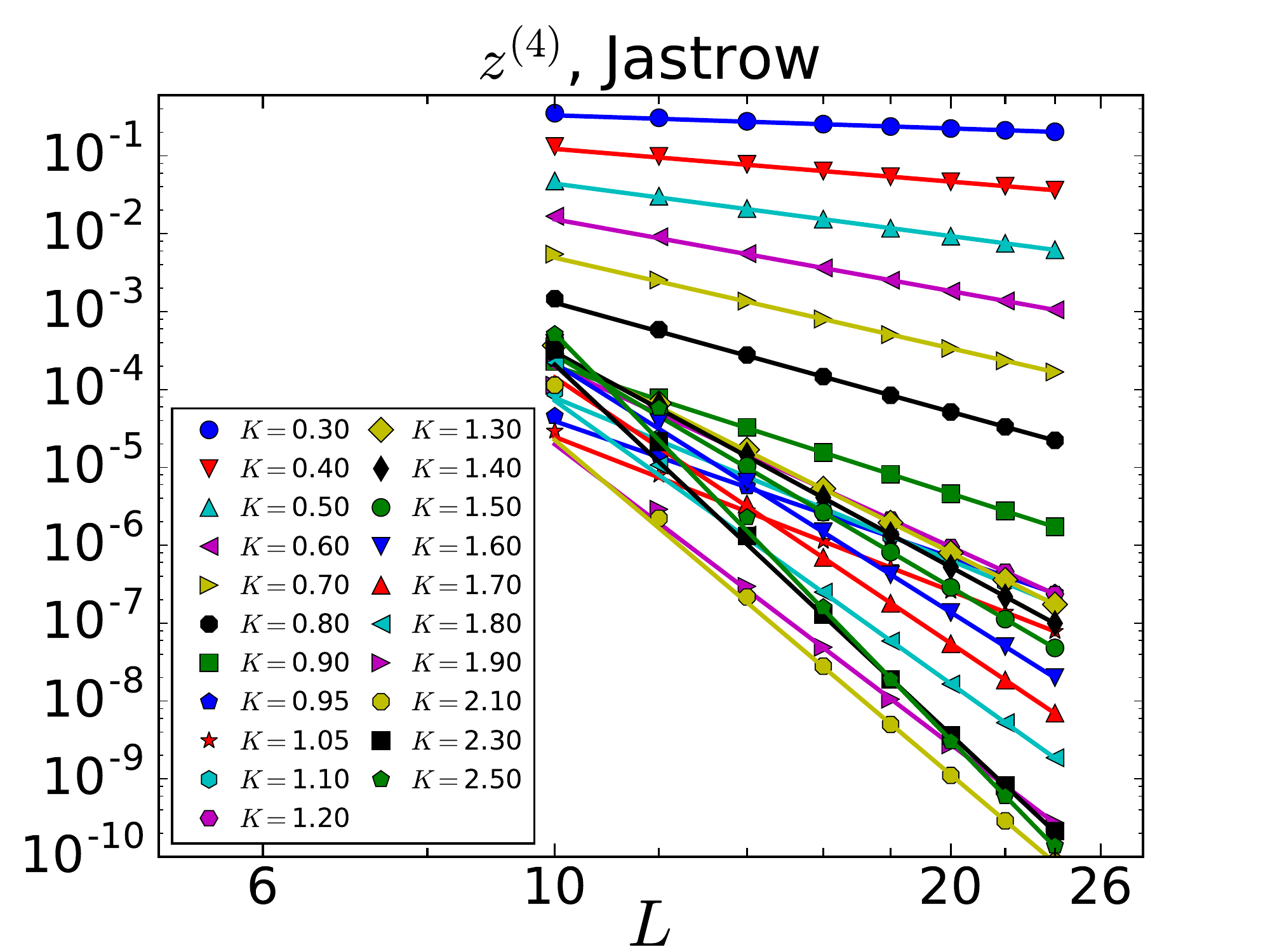}
 \includegraphics[width = 8cm]{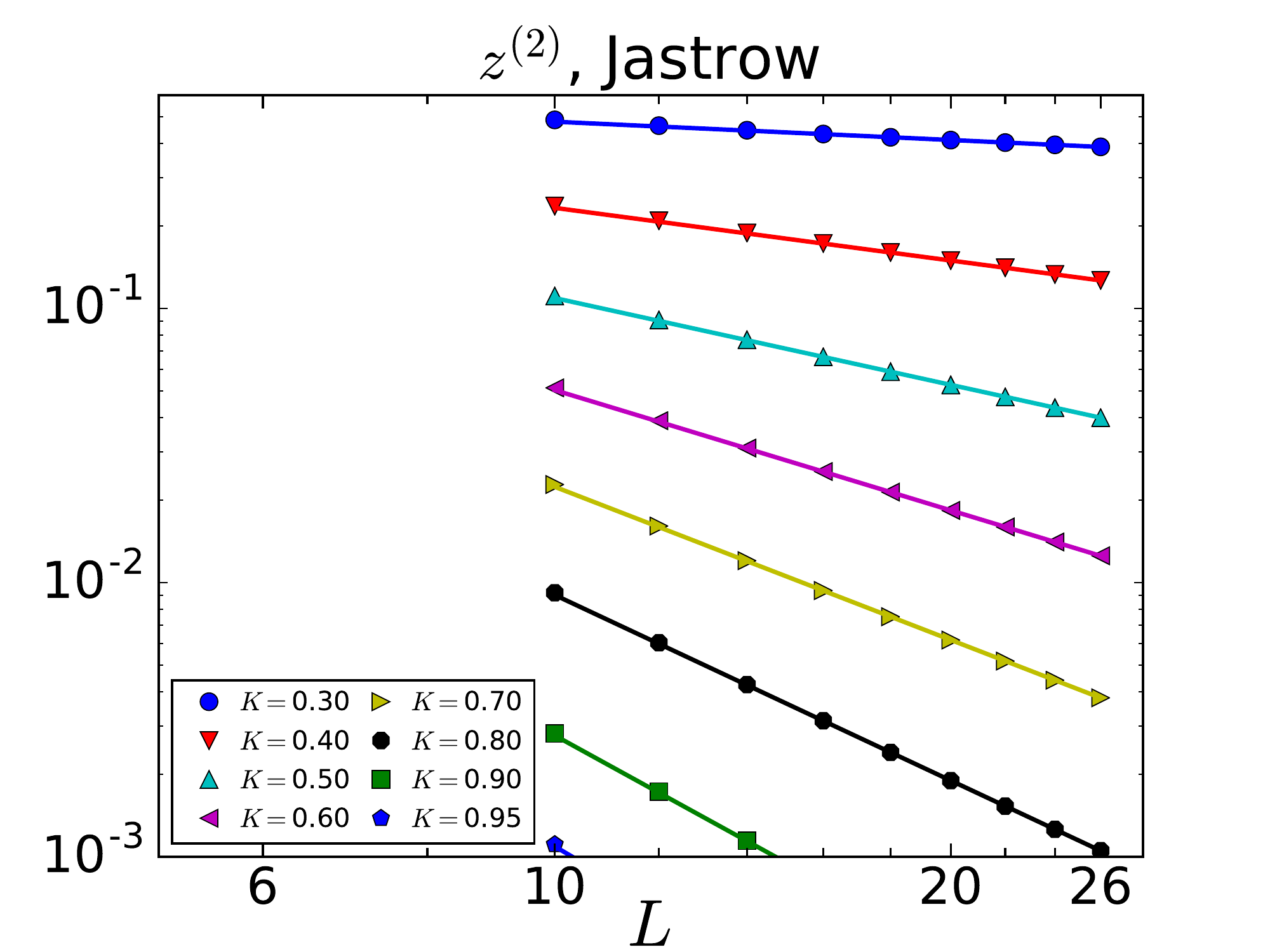}
 \includegraphics[width = 8cm]{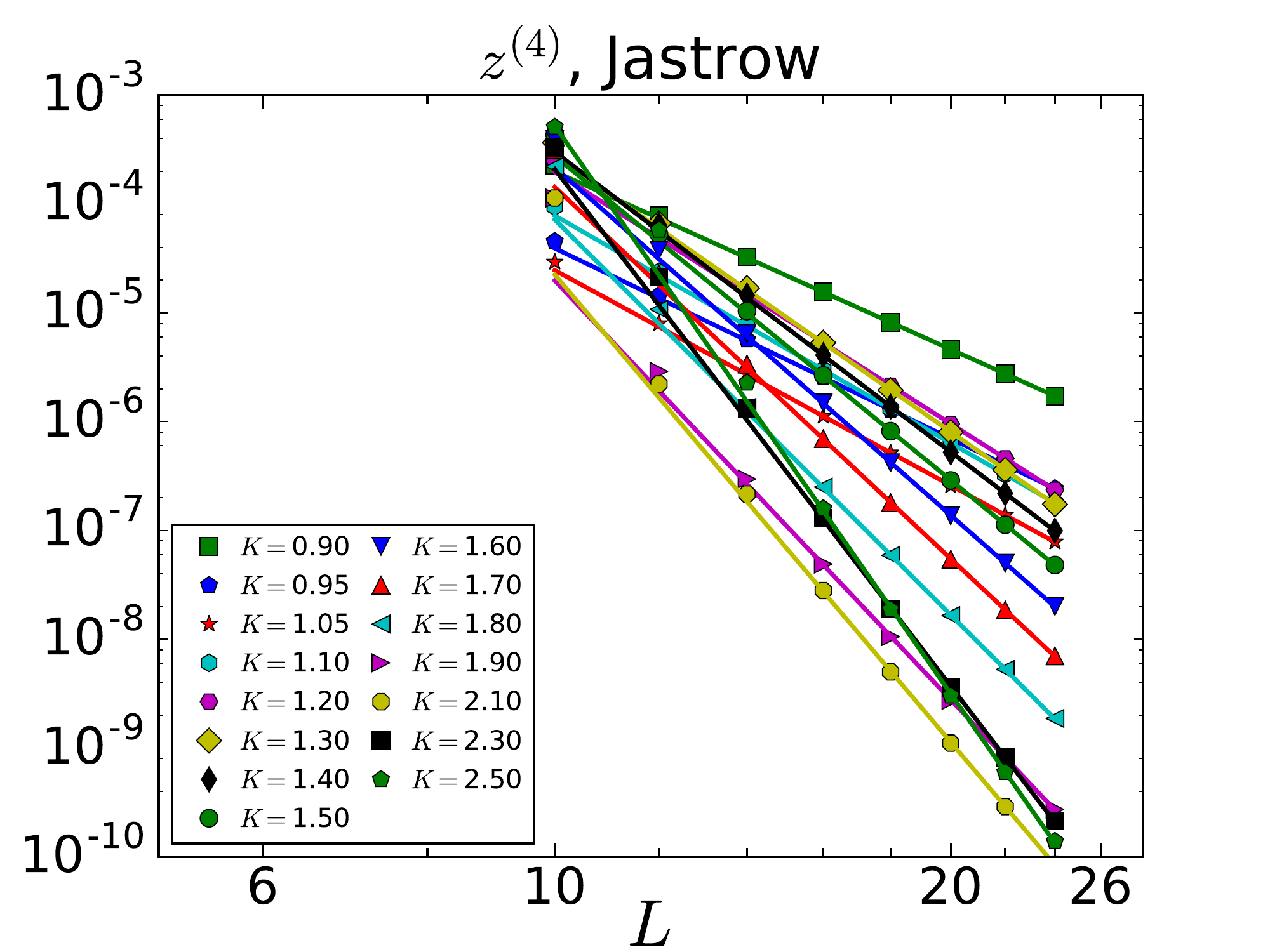}
 \includegraphics[width = 8cm]{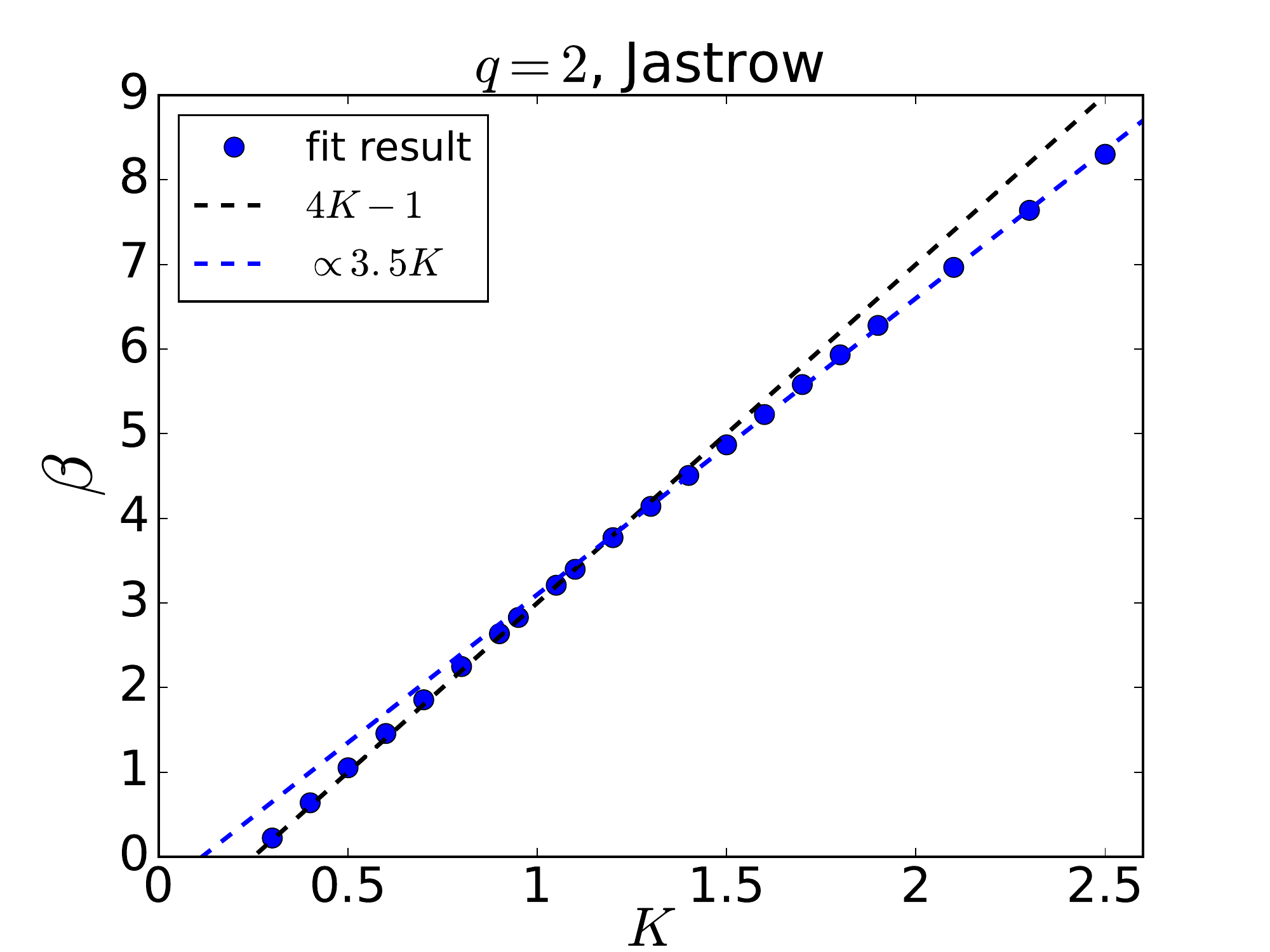}
 \includegraphics[width = 8cm]{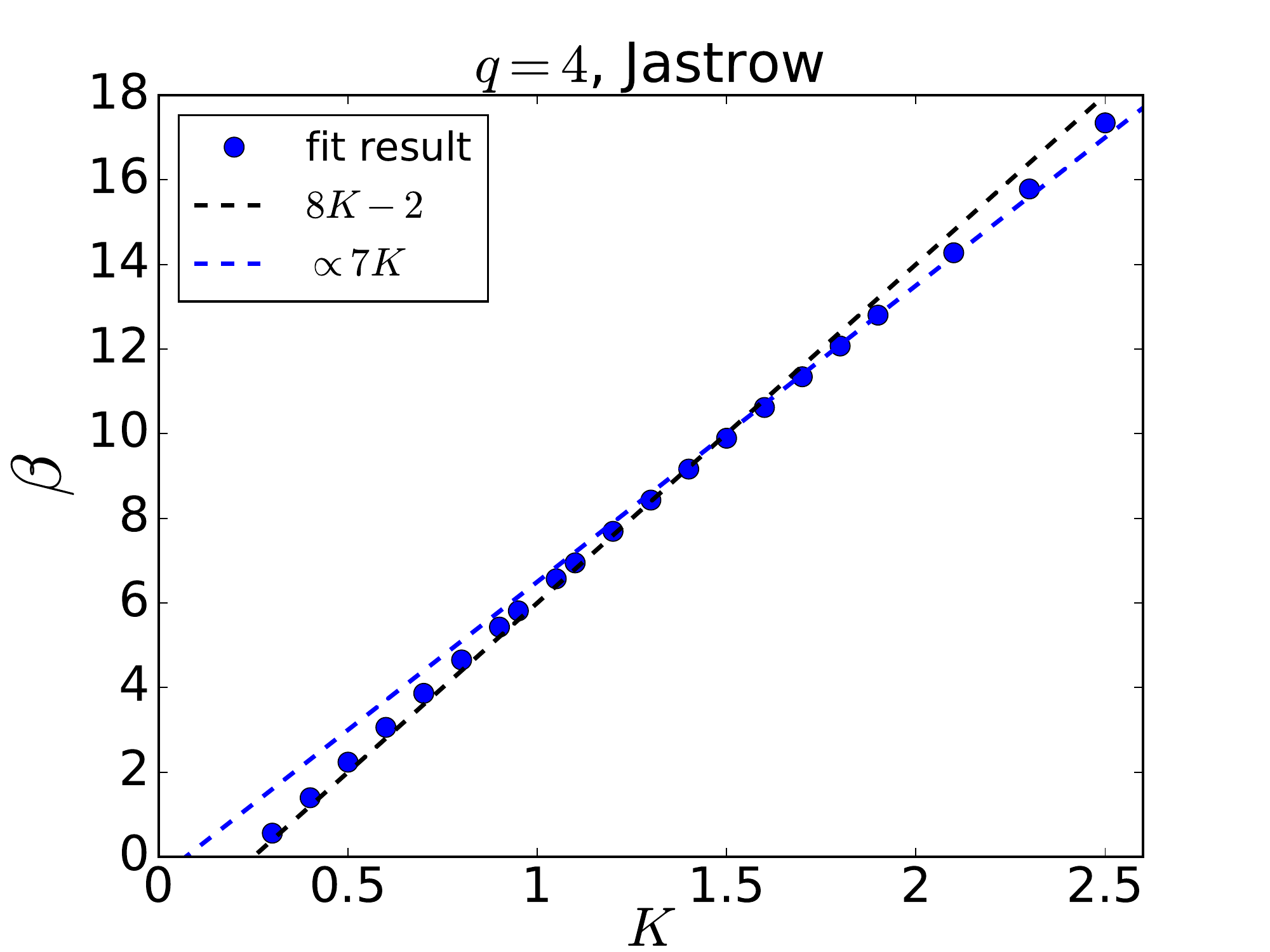}
 \caption{
 (top)
 Numerical results of $\z2$ (left) and $z^{(4)}$ (right) for the Gutzwiller-Jastrow wave function state~\eqref{Jastrowstate} with the system size $L$.
 The dots are the numerical data and the lines are numerical fits by a simple power-law $f(L) = a / L^\beta$ where $a$ and $\beta$ are fitting parameters.
 (middle) The closeups of the top panels.  
 (bottom) The power-law exponent $\beta$ obtained from the fitting.
  }
 \label{Jastrow_numerics}
\end{figure*}

The results of $\z2$ are shown in the top left and middle left panels of Fig.~\ref{Jastrow_numerics}, where one can see a clear power-law behavior of $\z2$ with $L$.
We also present the $K$-dependence of the exponent of the power-law $\beta$ in the bottom left panel of Fig.~\ref{Jastrow_numerics}.
For $K \lesssim 1.5$, it seems that $\beta = 4K-1$ explains the data well. However, for $K \gtrsim 1.5$, the slope of the $\beta$-$K$ curve becomes small and $\beta \propto 3.5K$ seems to fit the data.
Generally the finite size effect is strong for large positive $K$ (ferromagnetic-like critical regime) as one can see in the XXZ chain and \J1J2 XXZ model described in the previous subsections,
but in this case the difference between $K < 1.5$ and $K>1.5$ is not due to the finite size effect
because the power-law behavior is evident even for $K > 1.5$ within the accessible system size $L=26$
in Fig.~\ref{Jastrow_numerics}.
The results of $z^{(4)}$ are qualitatively the same as those of $\z2$, so the exponent seems
$\beta = 8K-2 $ for $K \lesssim 1.5$ and $\beta \propto 7K$ for $K \gtrsim 1.5$
(see the right column of Fig.~\ref{Jastrow_numerics}).
Thus, for $K \lesssim 1.5$ we conjecture that
\begin{equation}
 \beta(q) = q \left(2K - \frac{1}{2}\right),
\label{eq.beta.Jastrow}
\end{equation}
for an even integer $q$ in the Gutzwiller-Jastrow wave function.
This is consistent with the exact result~\eqref{HSresults} for the
SU(2) symmetric Haldane-Shastry model with $K=1/2$.

We note that when $K=1/4$ our numerical finding~\eqref{eq.beta.Jastrow}
gives $\beta=0$.
This corresponds to the phase transition in the state~\eqref{Jastrowstate} into the gapped wave function~\cite{KumanoPhDthesis} where $\zq$ has a finite values even in the thermodynamic limit.


\section{Conclusion and Discussion} 
\label{sec:discussion}

In this paper, we have studied analytically and numerically the polarization amplitude $\zq$ proposed by Resta~\cite{Resta} and modified by Aligia and Ortiz~\cite{Aligia99}, in the gapless critical TLL phase of the $S=1/2$ XXZ chain and its generalizations.
We found the power-law scaling~\eqref{eq.zqscaling}, which confirms Resta's proposal that the polarization amplitude can be used as an ``order parameter'' to distinguish insulators and conductors.
On the other hand, the exponent $\beta$ is different among several models, even when they are described by the TLL with the same Luttinger parameter $K$. 
Our numerical results suggest that, when $ K \lesssim 1.5$,
the exponent $\beta$ for $\zq$ with an even integer $q$ is $q(2K-1)$ for the XXZ chain, $q(2K)$ for the \J1J2 XXZ chain at the Gaussian point, and $q(2K-1/2)$ for the Gutzwiller-Jastrow wavefunction.

It is interesting to note that the exponent $\beta$ approaches to zero in the limit of $K \to 1/2$ in the standard XXZ chain, and of $K \to 1/4$ for the Gutzwiller-Jastrow wavefunction. These are precisely when a phase transition to a gapped phase takes place. This seems to be consistent with Resta's original proposal that $\zq \neq 0$ signals an insulator. Exactly at the Heisenberg antiferromagnetic point $\Delta=1$, where $K=1/2$, the system is a gapless conductor. It would contradict with $\beta=0$ if the simple power-law scaling is assumed. However, at the Heisenberg point, we expect a logarithmic correction which makes the simple power-law scaling~\eqref{eq.zqscaling} invalid. Presumably, at the Heisenberg antiferromagnetic point, $\zq$ would follow a logarithmic scaling and vanishes in the thermodynamic limit.

The observed ``non-universality'' of the exponent in TLLs is not too surprising by itself. Physical quantities are often controlled by irrelevant perturbations to the conformal field theory which represents the infrared fixed point. In fact, the XXZ chain is described by the ideal TLL (free boson field theory) perturbed by the Umklapp terms. In the \J1J2 chain at the Gaussian point, on the other hand, the leading Umklapp term is fine-tuned to zero. Thus the difference in the exponent suggests that the the leading Umklapp term indeed has an important effect on the amplitude. However, at present, we do not have a field-theory derivation of the observed numerical results.
In fact, even for the Haldane-Shastry model, in which all the Umklapp terms vanish, the prediction from the free boson field theory does not match the exact result.
It is also quite puzzling that the Haldane-Shastry model has the exponent $\beta(q)=q/2$ \emph{in between} that for the Heisenberg antiferromagnetic chain (XXZ chain with $\Delta=1$), $\beta(q)=0$, and the \J1J2 model with $\Delta=1$ at the Gaussian point, $\beta(q)=q$.
Suppression of the amplitude $\zq$ by eliminating the leading Umklapp term might explain the larger exponent for the \J1J2 model at the Gaussian point compared to the Heisenberg antiferromagnetic chain. However, similar argument would contradict with the smaller exponent for the Haldane-Shastry model where all the subleading Umklapp terms are supposed to be absent.
Our results indicate importance of the Umklapp terms but their exact role remains a mystery.
Moreover, we observe an apparent change in the behavior of $\beta$ across $K=1.5$ for all three states. We have no idea to explain this behavior, since there is no particular operator which becomes relevant $K=1.5$. 

The difficulty of the TLL description of $\zq$ is surprising, given the success of TLL approach in describing low-energy physics of one-dimensional quantum many body systems, in particular the $S=1/2$ chain. It may be that the amplitude $\zq$ is dominated by the short-distance/high-energy physics. However, since the state $U | \Psi_0\rangle$ is still a low-energy state~\cite{LSM61}, one would expect a field-theory description of the amplitude $\zq$. The possible field-theory description of the intriguing observations, including the change in the exponent around $K=1.5$, is left as a problem for the future.

\section*{Acknowledgment}
The authors thank Yohei Fuji, Shunsuke Furukawa,  Thierry Giamarchi, Chang-Tse Hsieh, Masaaki Nakamura, Kiyomi Okamoto, Ken Shiozaki, Haruki Watanabe, and Yuan Yao for valuable comments, and in particular Gil-Young Cho for a stimulating discussion which led to the present work.
RK and YON were supported by Advanced Leading Graduate Course for Photon Science (ALPS) of Japan Society for the Promotion of Science (JSPS).
This work was supported in part by from JSPS KAKENHI Grants Nos.\ JP16J01135 (Y.O.N.) and JP16K05469. (M.O.).

\appendix



\section{Naive calculation by Tomonaga-Luttinger liquid theory}
\label{app:TLL}

In this appendix, we present a field-theory approach to the
calculation of $\zq$, based on the TLL~\eqref{free boson action}.
 
A naive field-theoretical formulation of $\zq$ proceeds as follows. First, we define the bosonic field on the complex plane
\begin{equation}
\langle  \phi_{\mathrm{chiral}}(z)\phi_{\mathrm{chiral}}(z') \rangle =-K \log (z-z').
\label{free boson}
\end{equation}
The bosonic field for a periodic spin chain with length $L$ is defined on an infinite cylinder with circumference $L$.
We can calculate correlation functions on a cylinder, by performing the conformal transformation $z= \exp \left( {2\pi iw}/{L}\right)$ on a complex plane, to map correlation functions calculated on a complex plane to those on a cylinder.

For convenience, we fix the time variable $t=0$ and omit it. 
The Lagrangian density for the theory is
\begin{equation}
\mathcal{L}=\frac{1}{2\pi K} \left[\left( \partial_{\tau} \phi  \right)^{2}+\left( \partial_{x} \phi  \right)^{2}\right].
\label{CFTlagrangean}
\end{equation} 
With the perturbation $\cos\left(2\phi\right)$, the theory changes massive to massless
at $K=1/2$, which corresponds to the Heisenberg point. 
$0<K<1$ corresponds to the antiferromagnetic region, and $K >1$ corresponds to the ferromagnetic region.
The spin operator for $z$ direction is expected to be represented by
\begin{equation}
S^{z}_{j}=\frac{1}{2\pi } \partial_{x} \phi+\left( -1\right)^{j }\cos \left( \phi\right).
\end{equation}
By assuming the cancellation of the staggered terms, one can easily see the magnetization
of the chain corresponds to the topological sector for the free boson with periodic boundary condition
\begin{equation}
\phi (x+L,t)=\phi(x,t)+2\pi mL,
\end{equation}
where $m$ is the magnetization. 

Then, a naive expected form for the polarization represented by free boson on the infinite cylinder under the periodic boundary condition ($m=0$) is
\begin{align}
	\begin{split}
	\zq&=\left\langle \exp\left(\frac{2\pi qi}{L}\sum_{j=1}^L j\cdot \left(\frac{1}{2\pi}\partial_x\phi(j)+(-1)^j\cos\phi(j)\right)\right)\right\rangle \\
	&\approx \left\langle \exp\left(\frac{qi}{L}\int_0^L x\partial_x\phi dx\right) \exp\left(\frac{q\pi i}{L}\int_0^L (x\partial_x\cos\phi+\cos\phi) dx\right) \right\rangle \\
	&= \left\langle \exp\left(iq\phi(0) -\frac{iq}{L}\int^{L}_{0} \phi dx\right)\exp(iq\pi\cos\phi(0)) \right\rangle \\
	&= \left\langle \exp\left(iq\phi(0) -\frac{iq}{L}\int^{L}_{0} \phi dx\right)\sum_{n=0}^\infty\frac{(iq\pi\cos\phi(0))^n}{n!} \right\rangle,
	\end{split}
\end{align}
where we performed the partial integration in the third equality. We assume that one can switch the expectation and the integral in the following manner
\begin{equation}
\zq=\lim_{N\rightarrow\infty} \left\langle \exp \left( iq\phi(0) -\frac{iq}{N}\sum^{N-1}_{j=0} \phi \left( \frac{jL}{N}\right)\right) \sum_{n=0}^\infty\frac{(iq\pi\cos\phi(0))^n}{n!}        \right\rangle.
\label{multipoint function}
\end{equation}
Here, we see that $\zq$ reduces to a multi-point correlation function of vertex operators $V_\alpha(x):=\exp(i\alpha\phi(x))$ located on a ring characterized as $t=0$. We remark that the multi-point function of vertex operators $\left\langle\prod_{j}V_{\alpha_j}(x_j)\right\rangle$ becomes finite iff the conformal Ward identity $\sum_{j}\alpha_j=0$ is satisfied~\cite{YellowBook}. Then, by picking up terms that satisfy the conformal Ward identity from (\ref{multipoint function}), 
\begin{align}
\zq\propto \lim_{N\rightarrow\infty} \left\langle V_q(0)\prod_{j=0}^{N-1} V_{-q/N}\left( \frac{jL}{N}\right) \right\rangle.
\label{zqmul}
\end{align}
we can calculate such a quantity by transforming from an infinite cylinder to a complex plane, performing the conformal transformation $z= \exp \left( {2\pi iw}/{L}\right)$. The correlation function is transformed as
\begin{align}
\left\langle\prod_{j}V_{\alpha_j}(z_j)\right\rangle = \prod_{j}\left(\frac{dw_j}{dz_j}\right)^{2K\alpha_j^2}\cdot \left\langle\prod_{j}V_{\alpha_j}(w_j)\right\rangle.
\end{align}
Then, by identifying $z_j, \alpha_j$ as $z_j=\exp(2\pi i j/N)$ and
\begin{align}
\alpha_j=
\begin{cases}
{(N-1)q}/{N} & j=0 \\
{-q}/{N} & j\neq 0,
\end{cases}
\end{align}
(\ref{zqmul}) is rewritten as
\begin{align}
	\begin{split}
	\zq&=\lim_{N\rightarrow\infty}\prod_{j=0}^N\left(\frac{2\pi iz_j}{L}\right)^{2K\alpha_j^2}\cdot \left\langle\prod_{j}V_{\alpha_j}(z_j)\right\rangle \\
	&\propto\left(\frac{1}{L}\right)^{2q^2 K}\cdot\lim_{N\rightarrow\infty} \prod_{j=1}^{N-1}\lvert1-z_j\rvert^{-{q^2}/{N}}\prod_{i<j} \lvert z_i-z_j\rvert^{q^2/N^2} \\
	&=\left(\frac{1}{L}\right)^{2q^2 K}\exp\left(-q^2\int^\pi_0\log(2\sin x)dx\right)\exp\left(\frac{q^2}{2}\int^{\frac{\pi}{2}}_0\log(2\sin x)dx\right) \\
	&=\left(\frac{1}{L}\right)^{2q^2 K},
	\end{split}
\label{zqcft}
\end{align}
where we used  
\begin{align}
\int^{\frac{\pi}{2}}_0\log(\sin x)dx=-\frac{\pi}{2}\log2
\end{align}
 in the last equality. Moreover, one can obtain general results for the general topological sector corresponding to finite magnetization, by the following redefinition of the bosonic
field
\begin{align}
\phi(x+L)=\phi(x)+2\pi mL, \ 
& \phi'(x)=\phi(x)-2\pi mx, 
\label{topological sector}
\end{align}
with $m$ being magnetization. This $\phi'$ satisfies the periodic boundary condition, and we get the following result
\begin{equation}
\zq\propto e^{i\pi qmL}\left( 1/L\right) ^{2q^2K}.
\end{equation}
According to (\ref{zqcft}), the power law exponent $\beta(q)$ of $\zq$ satisfies $\beta(q) \propto q^2$, which reflects that the conformal dimension of the vertex operator $V_{\alpha q}$ becomes proportional to $q^2$. On the other hand, we will see in Section \ref{sec:weak}, \ref{sec:HS}, \ref{sec:num} that $\beta(q)\propto q$ for the ground state of XXZ model, $J_1$-$J_2$ model, and Gutzwiller-Jastrow wave function, which manifestly contradicts the result obtained by the above CFT-based calculation (\ref{zqcft}). Moreover, in (\ref{zqcft}) $\zq$ takes finite value for arbitrary integer $q$, which is wrong with the fact that $\zq$ is enforced to vanish by the lattice one-site translation symmetry when $q$ is odd, as observed by Aligia and Ortiz~\cite{Aligia99}. These conflicts may be resulting from the fact that this field-theoretic formulation does not respect the translation symmetry, which should be implemented as $\phi(x)\rightarrow\phi(x+1)$. For the expression (\ref{zqcft}), the translation about a unit cell transforms $\zq$ as
\begin{equation}
\zq \rightarrow e^{-2\pi iqm} \exp \left( iq \left( \phi(1)-\phi(0)\right)\right) \zq.
\end{equation}
This expression means $\zq$ is not invariant under one site translation. Although we do not know how to calculate correlation functions respecting the lattice translation symmetry for finite size systems, the commutation relation between $U$ and the translation operator $T$ (\ref{UT}), which enforces $\zq=\langle U^q \rangle$ to vanish, is successfully translated into the language of the field theory, in the thermodynamic limit $L\rightarrow\infty$~\cite{Gil17}. In the thermodynamic limit, the lattice translation $T$ becomes on-site symmetry because the lattice constant is brought to zero in this limit, and we can discuss the quantum anomaly with respect to $T$. In such a situation, it is pointed out~\cite{Gil17} that the commutation relation (\ref{UT}) is equivalent to the mixed quantum anomaly for $U(1)\times T$ symmetry, i.e., the phase ambiguity of the partition function under the large gauge transformation~\cite{Matsuo} represented as $U$, in the presence of the twist by $T$.

\section{Norm of the Gutzwiller-Jastrow wavefunction}
\label{app:norm}

Although the derivation and the result are well known~\cite{Kato09}, we will review calculations of the norm of the Gutzwiller-Jastrow wavefunction here, as these calculations will be useful also as a warming-up for the calculation
of the polarization.
First we observe
\begin{equation}
 |z_i - z_j |^4 = (z_i z_j)^{-2} (z_i - z_j)^4 .
\end{equation}
Thus 
\begin{equation}
\prod_{i<j} |z_i - z_j|^4 =
\prod_j {z_j}^{-2(M-1)} \prod_{i<j} (z_i - z_j)^4.
\end{equation}
Using~\cite{Sobczyk02}
\begin{equation}
\prod_{i<j} (z_i - z_j)^4 =
\det{\begin{pmatrix}
{z_k}^l, l {z_k}^{l-1}
\end{pmatrix}
}_{l=0, \ldots 2M-1}^{k=1,\ldots,M},   
\end{equation}
it follows that
\begin{align}
|\tilde{\Psi}_G|^2 &= 
\prod_j {z_j}^{-2(M-1)} \prod_{i<j} (z_i - z_j)^4  \notag \\ 
&=
\prod_j {z_j}^{-2(M-1)} 
\det{\left( {z_k}^l, l {z_k}^{l-1} \right)} \notag \\
&=
\det{\left( {z_k}^{l-M+\frac{1}{2}}, l {z_k}^{l-M+\frac{3}{2}-1}\right)}
\notag \\
&=
\det{\left( {z_k}^{l-M+\frac{1}{2}}, 
(l-M+\frac{1}{2}) {z_k}^{l-M+\frac{1}{2}}
\right)}
\notag \\
&=
\det{\left(
e^{i p_l \theta_k}, p_l e^{i p_k \theta_k}
\right)
} ,
\end{align}
where
\begin{equation}
 p_l = -M + \frac{1}{2}+ l = 
-M + \frac{1}{2}, -M + \frac{3}{2}, \ldots M-\frac{1}{2} .
\label{eq.range_pl}
\end{equation}
From this range of $p_l$, 
\begin{equation}
  -2(M-1) \leq p_l + p_k \leq 2(M-1),
\label{eq.range_sum_pj}
\end{equation}
for $k \neq l$.

Now we expand the determinant as
\begin{equation}
 \sum_P \epsilon_P
 p_{P2} e^{i(p_{P1} + p_{P2}) \theta_1} 
 p_{P4} e^{i(p_{P3} + p_{P4}) \theta_2} 
 \ldots
 p_{P(2M)} e^{i(p_{P(2M-1)} + p_{P(2M)}) \theta_M} ,
\end{equation}
where $P$ denotes a permutation.
Combining $P$ with $P_{2j-1,2j} P$ ($P_{2j-1,2j}$ denotes the transposition between $2j-1$ and $2j$),
\begin{align}
|\tilde{\Psi}_G|^2 = 
\left(\frac{1}{2} \right)^M
 \sum_P \epsilon_P &
 (p_{P2} - p_{P1}) e^{i(p_{P1} + p_{P2}) \theta_1} 
\notag \\
& (p_{P4} - p_{P3}) e^{i(p_{P3} + p_{P4}) \theta_2} 
\notag \\
& \ldots
 (p_{P(2M)}-p_{P(2M-1)}) e^{i(p_{P(2M-1)} + p_{P(2M)}) \theta_M} .
\label{eq.norm_expand}
\end{align}
Because of Eq.~\eqref{eq.range_sum_pj}, we find 
\begin{equation}
 \sum_{\theta_j} e^{i(p_{P(2j-1)} + p_{P(2j)}) \theta_j} 
 = 
\begin{cases}
L & (p_{P(2j-1)} + p_{P(2j)} = 0), \\
0 & \mbox{otherwise}
\end{cases} 
\end{equation}
Thus, the non-vanishing contributions to Eq.~\eqref{eq.norm_expand}
are limited to $P$ which satisfy
\begin{equation}
 p_{P(2j-1)} + p_{P(2j)} = 0 .
\label{eq.sump_0}
\end{equation}
Within the set~\eqref{eq.range_pl}, there are
$M$ distinct pairs of $p_j$'s
\begin{equation}
(p_1, p_{2M}), (p_2, p_{2M-1}), (p_3, p_{2M-2}), \ldots
(p_M,p_{M+1})
\label{eq.pairs_p}
\end{equation}
that satisfy Eq.~\eqref{eq.sump_0}.
Since all $p_j$'s are different,  we need to have exactly $M$
distinct pairs that satisfy Eq.~\eqref{eq.sump_0} to appear
in Eq.~\eqref{eq.norm_expand}.
For the pair $(p_j,p_{2M-j})$, we have
\begin{equation}
 p_{2M-j} - p_j = 2M-2j+1 .
\label{eq.diff_p}
\end{equation}
Since there are $M!$ ways of reordering the pairs, and
the product of the difference~\eqref{eq.diff_p} is
$(2M-1)!!$, the norm is given as Eq.~\eqref{eq.norm}.


\section{Details on numerical calculations of the \J1J2 XXZ model 
\label{app: J1J2 Gaussian} }
Here we explain how to determine the value of $J_{2,G}(\Delta)$ in the model~\eqref{J1J2ham} which corresponds to the Gaussian fixed point and the corresponding Luttinger parameter $K$.
To determine $J_{2,G}(\Delta)$
we employ the level spectroscopy method invented by Okamoto and Nomura~\cite{Nomura-Okamoto94, Nomura95} which
wisely combines the knowledge of the conformal field theory and actual numerical data with the finite system size $L$.

Since there are several symmetries in the model, we can assign quantum numbers to each energy eigenstate.
 We consider the total magnetization $m := \sum_{i=1}^L S_i^z$, the momentum $k$, and the parity $P$ under $S^z$-inversion. 
When the system size is a multiple of 4, the ground state of~\eqref{J1J2ham} is in the sector of 
$(m, k, P)=(0, 0, 1)$.
We introduce the dimer excitation state and the N\'eel excitation state,
which are the ground states of the sector of $(m, k, P) = (0, \pi, 1)$ and $(m, k, P) = (0, \pi, -1)$, respectively.
At the Gaussian fixed point these two excitation energies coincide~\cite{Nomura-Okamoto94}, so it is possible to estimate the $J_{2,G}(\Delta)$ by tracking the excitation energies with varying $J_2$ while fixing $\Delta$.

Actual numerical determination proceeds as follows:
(1) by varying $J_2$ with fixed $\Delta$ at the finite system size $L$,
we collect the value of the Gaussian fixed point $J_{2,c}(\Delta; L)$ by using the exact diagonalization.
(2) then we perform $1/L^2$-scaling to the $J_{2,c}(\Delta; L)$,
$J_{2,c}(\Delta; L) = J_{2,c}(\Delta) + \mathrm{const.} \times L^{-2}$,
to obtain the value in the thermodynamics limit, $J_{2,c}(\Delta) = \lim_{L \to \infty} J_{2,c}(\Delta; L)$.

As for the Luttinger parameter $K$ corresponding to numerically obtained $J_{2,G}(\Delta)$,
we utilize the level spacing between the excited states which correspond to the primary fields of the conformal field theory. We focus on two primary excited states: the $i \partial \phi$ state whose scaling dimension is 1 and the $e^{i\theta}$ state whose scaling dimension is $1/4K$~\cite{YellowBook}.
When the system size $L$ is a multiple of 4, the ground state of the Hamiltonian~\eqref{J1J2ham} is in the sector of vanishing total magnetization $m = 0$ and momentum $k=0$ whereas
the primary state $e^{i\theta}$  is the ground state of the sector of
$m = 1 $ and $k=\pi$. The $ i\partial \phi $ state is the ground state of the sector of $m =0$ and $k=2\pi/L$.
For each value of $\Delta$ and $J_{2,G}(\Delta)$, the ratio between excitation energies of the $i \partial \phi$ state and the $e^{i\theta}$ state in finite size system $L$ is calculated
and the value of the ratio in the thermodynamic limit $L \to \infty$ is obtained by extrapolating it with $1/L^2$-scaling. Then we identify it as $1/4K$ and determine $K$.


\end{document}